\documentclass[preprint2]{aastex631}

\usepackage{graphicx}
\newcommand{\y}[1]{\ensuremath{y_{#1}}\xspace}

\newcommand{\m}{\ensuremath{\,{\rm m}}\xspace}

\newcommand{\K}{\ensuremath{\,{\rm K}}\xspace}

\newcommand{\msun}{\ensuremath{{\rm \,M_\odot}}\xspace}     % solar mass
\newcommand{\h}[1]{\ensuremath{^{#1}{\rm H}}\xspace}

\newcommand{\hii}{{\rm H\,{\small II}}\xspace}

\usepackage[utf8]{inputenc}
\usepackage{xspace}
\usepackage{natbib}

\usepackage{url}

\def\refreport#1{#1}

\shorttitle{Protostellar cores in Sgr B2}
\shortauthors{Budaiev et al.}

\accepted{September 15, 2023}
\submitjournal{ApJ}

\begin{document}

\title{Protostellar cores in Sagittarius B2 N and M}

\correspondingauthor{Nazar Budaiev}
\email{nbudaiev@ufl.edu}

\author[0000-0002-0533-8575]{Nazar Budaiev}
\affiliation{Department of Astronomy, University of Florida, P.O. Box 112055, Gainesville, FL 32611-2055, USA}

\author[0000-0001-6431-9633]{Adam Ginsburg}
\affiliation{Department of Astronomy, University of Florida, P.O. Box 112055, Gainesville, FL 32611-2055, USA}

\author[0000-0003-0416-4830]{Desmond Jeff}
\affiliation{Department of Astronomy, University of Florida, P.O. Box 112055, Gainesville, FL 32611-2055, USA}

\author[0000-0002-2542-7743]{Ciriaco Goddi} 
\affil{
Universidade de S{\~a}o Paulo, Instituto de Astronomia, Geof{\`i}sica e Ci{\^e}ncias Atmosf{\'e}ricas, Departamento de Astronomia, S{\~a}o Paulo, SP 05508-090, Brazil}
\affiliation{Dipartimento di Fisica, Universit\'a degli Studi di Cagliari, SP Monserrato-Sestu km 0.7, I-09042 Monserrato,  Italy}
\affiliation{INAF - Osservatorio Astronomico di Cagliari, via della Scienza 5, I-09047 Selargius (CA), Italy}
\affil{
INFN, Sezione di Cagliari, Cittadella Univ., I-09042 Monserrato (CA), Italy}

\author[0000-0002-5927-2049]{Fanyi Meng}
\affiliation{University of Chinese Academy of Sciences, Beijing 100049, PR China}
\affiliation{I. Physikalisches Institut der Universit\"at zu K\"oln,
Z{\"u}lpicher Str. 77, 50937 K\"oln, Germany}

\author[0000-0002-3078-9482]{\'Alvaro S\'anchez-Monge}
\affiliation{Institut de Ci\`encies de l'Espai (ICE, CSIC), Can Magrans s/n, E-08193, Bellaterra, Barcelona, Spain}
\affiliation{Institut d'Estudis Espacials de Catalunya (IEEC), Barcelona, Spain}

\author[0000-0003-2141-5689]{Peter Schilke}
\affiliation{I. Physikalisches Institut der Universit\"at zu K\"oln,
Z{\"u}lpicher Str. 77, 50937 K\"oln, Germany}

\author[0000-0002-1730-8832]{Anika Schmiedeke}
\affiliation{Green Bank Observatory, PO Box 2, Green Bank, WV 24944, USA}

\author[0000-0003-2968-5333]{Taehwa Yoo}
\affiliation{Department of Astronomy, University of Florida, P.O. Box 112055, Gainesville, FL 32611-2055, USA}

\begin{abstract}
We present 500 AU and 700 AU resolution 1 mm and 3 mm ALMA observations, respectively, of protostellar cores in protoclusters Sagittarius B2 (Sgr B2) North (N) and Main (M), parts of the most actively star-forming cloud in our Galaxy.
Previous lower resolution (5000 AU) 3 mm observations of this region detected $\sim$150 sources inferred to be young stellar objects (YSOs) with $M>8\msun$. With a tenfold increase in resolution, we detect 371 sources at 3 mm and \refreport{218} sources in the smaller field of view at 1 mm. 
The sources seen at low resolution are observed to fragment into an average of two objects. About a third of the observed sources fragment.
Most of the sources we report are marginally resolved and are at least partially optically thick. 
We determine that the observed sources are most consistent with Stage 0/I YSOs, i.e., rotationally supported disks with an active protostar and an envelope, that are warmer than those observed in the solar neighborhood.
We report source-counting-based inferred stellar mass and the star formation rate of the cloud: \refreport{2800\msun, 0.0038\msun yr$^{-1}$ for Sgr B2 N and 6900\msun, 0.0093\msun yr$^{-1}$ for Sgr B2 M respectively.}
\end{abstract}

\section{Introduction} \label{sec:intro}
Many stars in our Galaxy were formed around cosmic noon \citep[$z\sim2$,][]{Madau2014} under conditions rarely seen today.
To study the conditions under which an average Galactic star formed, we need to identify star-forming regions which are analogous to those present during cosmic noon.
One such local environment analogue is present in the high-density, turbulent, and overall extreme Central Molecular Zone \refreport{\citep[CMZ;][]{Henshaw2022}}. This is the only environment with such conditions where we can resolve disk scales \refreport{\citep[$\sim$100s AU][]{Lu2022}}. 

The distribution of stellar masses, the initial mass function (IMF), is one outcome of star formation that may change under extreme gas conditions.
In young star clusters near the center of the Galaxy, such as the Arches and Quintuplet, there is evidence of a top-heavy IMF \citep[][]{Hosek2019a, Gallego-Calvente2022}. 
By peering into early stages of star formation in the CMZ, we will investigate whether the IMF varies systematically with local environment \refreport{\citep{Li2023}} and is thus shallower in the early universe \refreport{\citep{Zhang2018}}. In this work, we examine the most actively star-forming cloud in the CMZ -- Sagittarius B2 (Sgr B2). \refreport{Through investigating dust structures at $<1000$ AU scales, we can probe the core mass function -- the precursor to the IMF.}

The Galactic Center is located at a distance of $8.127\pm0.031$ kpc \citep{GRAVITYCollaboration2018} and Sgr B2 is $\sim$100 pc away from the Galactic Center in projection. Throughout this paper, we assume a distance to Sgr B2 of 8.4 kpc to remain consistent with the similar work by \cite{Ginsburg2018}. The molecular cloud is split into several smaller, concentrated regions of star formation: North (N), Main (M), South (S), and Deep South (DS). 
A large fraction ($\sim$25-50\%) of the current star formation in this cloud occurs in the N and M regions, with many stars forming in bound clusters \citep[][]{Ginsburg2018, Schmiedeke2016}. Half of the total star formation in the CMZ occurs in Sgr B2, despite containing $\sim$10\% of the mass ($8 \times 10^6$\msun) and $<1\%$ of the volume  \citep[$5\times10^4$ pc$^3$;][]{Ginsburg2018, Barnes2017, Schmiedeke2016}.

Sgr B2 appears to be forming stars preferentially in dense, bound clusters \citep{Ginsburg2018, Ginsburg2018CFE}. Since other Galactic Center star clusters have excesses of high-mass stars \citep[][]{Hosek2019a, Gallego-Calvente2022}, Sgr B2 is likely also forming extra massive stars. 3500 AU resolution observations towards Sgr B2 N and M suggest the presence of an excess of massive stars in the region \citep{Sanchez-Monge2017}.
If this excess is present, it indicates a systematic variation of the stellar IMF with gas conditions, with a shallower IMF under conditions prevalent in galaxies with rapid star formation.

We report follow-up ALMA observations to ALMA project 2013.1.00269.S, which covered a $15\times15$ pc area around the Sgr B2 complex at 3 mm \citep{Ginsburg2018} with $\sim$5000 AU resolution. Our data covers Sgr B2 N and M at 3 mm and at 1 mm. We establish the possible nature and evolutionary stage of the sources. We place lower limits on cores' masses in the attempt to understand whether the \refreport{core mass function} is top-heavy in young cluster and varies with the surrounding conditions. 

In Section \ref{sec:data} we describe the observations, imaging, and self-calibration procedures. We then explain the source extraction approach in Section \ref{sec:analysis}, make a comparison with previous low-resolution observations, and explain the way the spectral index was calculated between 3 mm and 1 mm. In Section \ref{sec:disc} we show that our sources are consistent with Stage 0/I YSOs by ruling out other possible options. We also present calculations of the inferred stellar mass of the cloud and the star formation rate and present the multiplicity function. The paper is summarized in Section \ref{sec:sum}. In the appendices we \refreport{quantify the line contribution to the continuum images}, discuss the details of cleaning and self-calibration, and present the radial core distribution for each protocluster.

\section{Observations and data reduction} \label{sec:data}

This paper utilizes the data from ALMA Cycle 4 project 2016.1.00550.S. The observations consist of two pointings centered on Sgr B2 N (ICRS 17:47:19.925; -28:22:18.608) and Sgr B2 M (ICRS 17:47:20.174; -28:23:04.788) with the 12m antennas. Both locations were observed in Band 3 and Band 6 with resolution of 0.096" $\times$ 0.073" ($\sim$700 AU) and 0.083" $\times$ 0.038" ($\sim$500 AU), respectively. 
The pointings cover $6.5\times4.5$ pc at 3 mm and roughly $1.75\times3.75$ pc at 1 mm out to primary beam level of 0.1.
The spectral setup for each Band consists of 4 spectral windows with 1920 channels each and channel frequency width of 976.6 kHz (1.875 GHz bandwidth), which corresponds to spectral resolution of $\sim$1.3 km/s at 226 GHz and $\sim$3.2 km/s at 92 GHz. The central spectral window frequencies are 217.84, 219.79, 231.80, and 233.67 GHz for Band 6 and 85.47, 87.37, 97.42, and 99.42 GHz for Band 3.

\begin{table*}[th]
\caption{Summary of the observations.}
\label{tab:observations}
\begin{tabular}{ccccc}
\hline
\hline
 Central frequency & Date        & $t_{science}$  & Baseline length range & \# of antennas \\
  GHz              &             &  seconds        &  meters               &             \\ \hline 
92.45              & 18-Sep-2017 & 2642            & 41-12100              &  43           \\ %Band 3, 1
92.45              & 18-Sep-2017 & 2644            & 41-12100              &  43           \\ %Band 3, 2
225.76             & 07-Sep-2017 & 2977            & 41-7600               &  46           \\ %Band 6, 1
225.76             & 16-Sep-2017 & 2973            & 41-12100              &  46         \\ \hline %Band 6, 2 

\end{tabular}
\end{table*}

The data were imaged using CASA version 5.7.0-134.el7 {\tt tclean} with robust parameter set to 0.5 \citep{McMullin2007}. \refreport{We found that for the majority of sources the line contribution was insignificant and thus no line subtraction or flagging was performed. We measure the contribution of emission and absorption lines in Appendix \ref{sec:line_sub}.}
\refreport{Each of the four pointings were imaged separately}. 
Despite the overlap between the Band 3 pointings, we used the standard gridder instead of the mosaic gridder. \refreport{A detailed explanation of the imaging choices is presented in Appendix \ref{sec:imaging}}.

The initial cleaned images contained many imaging artifacts, such as radial streaks, circular patterns, and parallel ``wavy" streaks. Therefore, we decided to self-calibrate the data to improve the quality of the final image. \refreport{We made sure that the phase self-calibration does not introduce spatial shifts and amplitude self-calibration does not change the flux erroneously}. A detailed explanation of the tested approaches is presented in Appendix \ref{sec:testing}.
In summary, our standard self-calibration procedure consisted of the following steps:
\begin{enumerate}
    \item Use one of the fully calibrated images from prior testing to estimate the lowest noise in the image. We measured the RMS in a box around an emission-free region close to the center of the image. 
    \item Use the same previously cleaned image to create a ``strict" contour mask at six-$\sigma$ threshold.
    \item Using the same image, draw a ``broad" mask that covers all regions where signal is present, even if it includes some artifacts or noise.
    \item Clean the image with the \texttt{tclean} threshold parameter set to six times the noise from step 1, and using the ``strict" mask from step 2. 
    \item Complete a phase self-calibration with \texttt{solint} set to \texttt{inf}.
    \item Repeat the clean from step 4.
    \item Self-calibrate the image with \texttt{solint} set to \texttt{int}.
    \item Clean the image with the \texttt{tclean} threshold parameter set to three times the noise from step 1 and using the ``broad" mask from step 3.
    \item Complete a round of phase and amplitude self-calibration with \texttt{solint} set to \texttt{inf}.
    \item Repeat the clean from step 8.
\end{enumerate}

\begin{figure*}[ht]
    \centering
    \includegraphics[width=\textwidth]{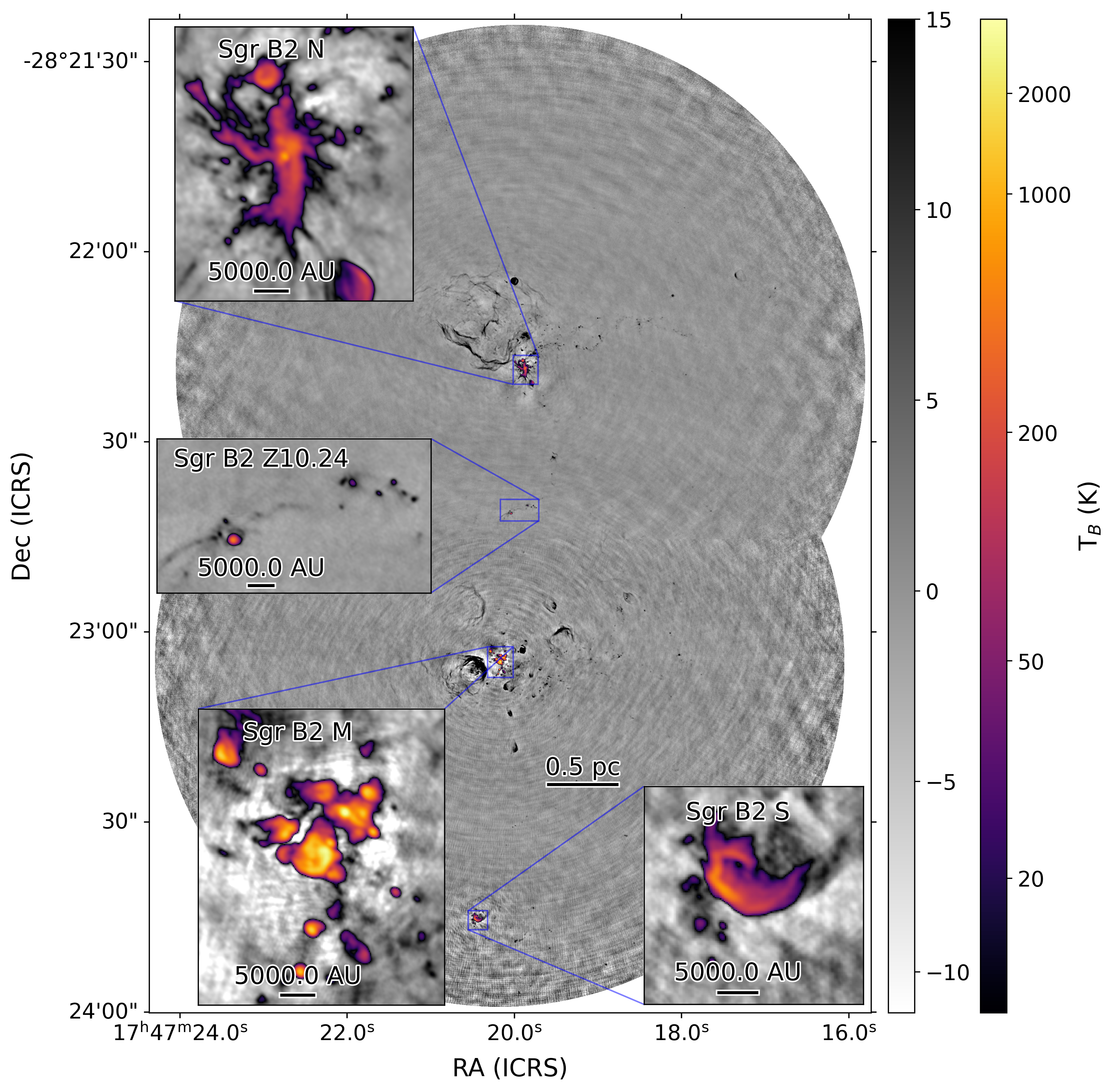}
    \caption{Band 3 (3 mm) continuum image of Sagittarius B2 N and M. The two pointings were imaged separately and then the final images were cut and overlaid to not overlap. The images were subsequently stitched together; the pixels in the central zoom-in come from only the N pointing. The insets show the high-density regions of Sgr B2 N and M, as well as Sgr B2 S and a region around \hii region Sgr B2 Z10.24. }
    \label{fig:B3}
\end{figure*}

\begin{figure*}[ht]
    \centering
    \includegraphics[width=\textwidth]{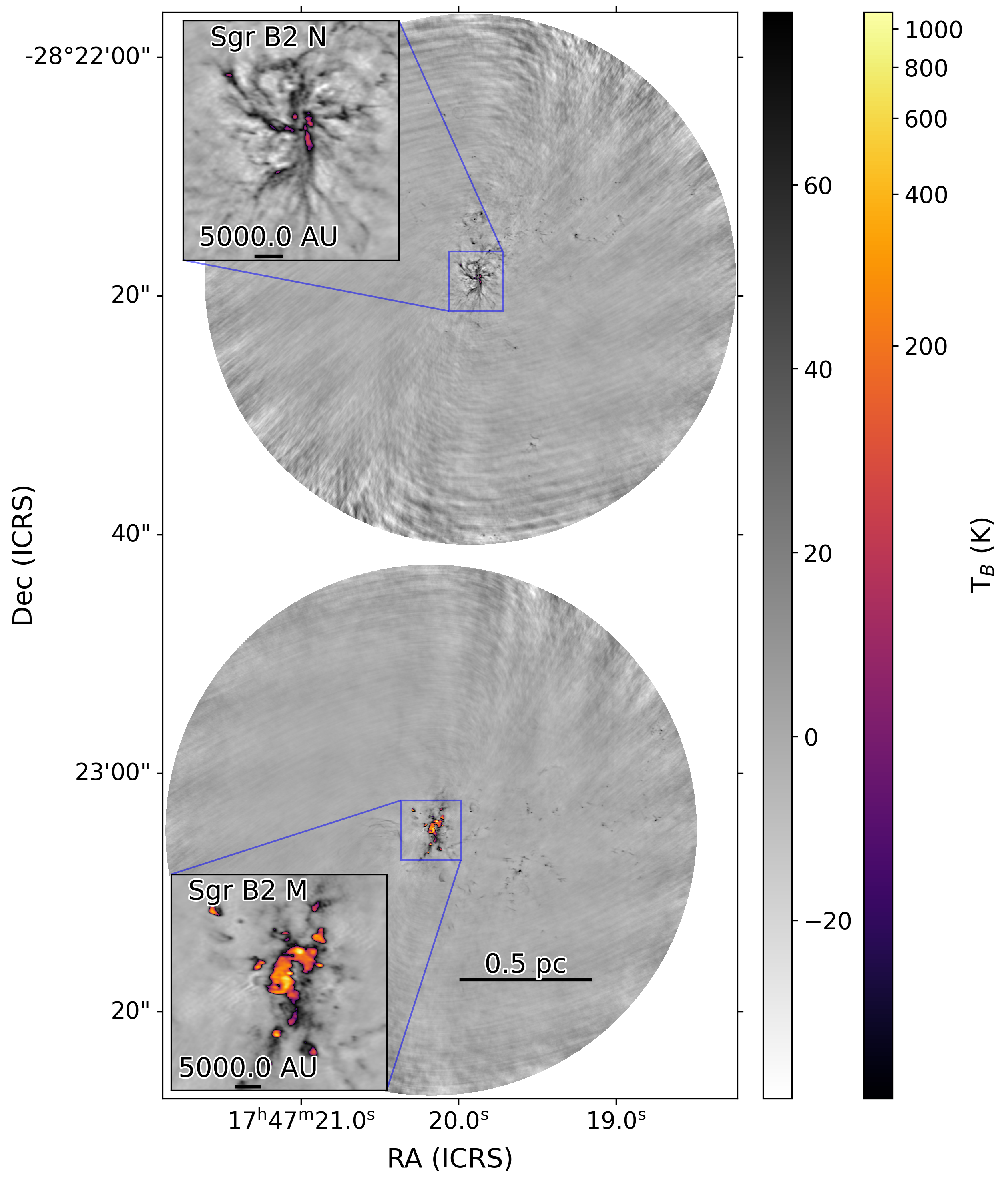}
    \caption{Band 6 (1 mm) continuum image of Sagittarius B2 N and M. Sgr B2 M appears to be hotter than Sgr B2 N. Due to the smaller field of view, \hii regions Sgr B2 S and Z10.24 that are visible in Band 3 are not present here.}
    \label{fig:B6}
\end{figure*}

However, some of the fields required further tuning to achieve better results. The Sgr B2 N Band 6 pointing suffered from divergence at the selected threshold values. Thus, we opted to increase the cleaning threshold up to ten times the noise until the amplitude self-calibration step was completed. After, we cleaned to three times the noise and added an extra amplitude and phase self-calibration step with \texttt{solint} set to 15 s, which we found to produce the best results. The extra self-calibration step improved the image significantly. Running this extra self-calibration step on the other pointings did not reduce the noise in the images.
The Sgr B2 N Band 3 pointing did not benefit from the amplitude self-calibration step and introduced additional minor artifacts. Thus, we used the pre-amplitude-self-calibration image for our analysis. The final images for each Band are shown in Figures \ref{fig:B3} and \ref{fig:B6}. The FITS files of the self-calibrated data were published in a Zenodo repository  at \href{https://doi.org/10.5281/zenodo.8377186}{doi:10.5281/zenodo.8377186}.

The final average RMS in an emission-free region close to the center the image for each of the pointings is 0.033 mJy for N Band 3, 0.05 mJy for M Band 3, 0.35 mJy for N Band 6, and 0.25 mJy for M Band 6. The RMS measurements for each calibration step are presented in Table \ref{tab:RMS}.

\section{Analysis} \label{sec:analysis}
\subsection{Source extraction} \label{sec:source_extraction}
We strove to make the catalog as reproducible as possible, while minimizing the number of false positives.
Automating source extraction presents numerous problems in a region like Sgr B2. 
The images have high dynamic range ($\sim$5000), contain complex extended structures, and like all interferometric single-pointing data, have radially varying noise. Combining this with the irregular structure of the observed objects makes fully automated source detection inefficient.
Therefore, we utilized the automated source-finding dendrogram algorithm \citep{Rosolowsky2008} augmented with a by-hand approach described below.

\refreport{The imaged areas of Band 3 fields have a slight overlap. Since the noise is not uniform and has structure depending on the location relative to the center of the pointing, we crop the overlapping parts instead of weighted averaging}.
The cut was made close to equidistant from the centers of the pointing, while not breaking up any regions with signal -- just south of (below) \hii region Sgr B2 Z10.24. Band 6 images do not overlap and thus do not require cropping.
We divide each primary-beam corrected image in 30 concentric annuli \refreport{out to the edge of the image, primary beam response of 0.1,} and a circle at the center. \refreport{The width of the concentric annuli is $\sim$$7.8$$''$ and $1.1''$ for Band 3 and Band 6 respectively}.
We then use 7-sigma clipping to remove most of the signal from each of the annuli and calculate the standard deviation for each region. This noise measurement is then used as \texttt{astrodendro} input parameters. 

The central part of the pointings are signal-dominated, which, even after sigma clipping, results in a higher measured noise.
The calculated noise as a function of radius rapidly drops beyond the central clusters and then gradually increases due to primary beam attenuation. We replace the estimated noise of the central part of the pointing with the lowest calculated value in the whole image such that the noise never decreases with radius.

Then, we perform source detection on the primary-beam-corrected images using the \texttt{astrodendro} source-finding algorithm. \refreport{Because of the extremely bright emission at the center of the pointings, the central region of the image is dominated by artifacts. Thus, the noise is not uniform in the non-primary-beam-corrected images.}
We generate a separate dendrogram for each concentric annulus using noise estimation for input parameters. We run the dendrogram on the full image and then exclude any sources with centers outside of the annulus. 
We set \textit{min\_value} to be equal to four times the noise estimate, \textit{min\_delta} to be 0.7 times the estimated noise, and \textit{min\_npix} to be 3 pixels.\footnote{We tested a variety of combination of number of concentric annuli (10, 15, 20, 30), \textit{min\_value} (3$\sigma$, 4$\sigma$, 5$\sigma$), \textit{min\_delta} (0.5, 0.7, 1.0), and \textit{min\_npix} (2, 3, 5). We inspected each combination by hand to find a compromise between reducing the number of false positives due to artifacts and noise and increasing the number of likely positives.} 
The final source extraction thresholds are $\sim$0.15 mJy at 3 mm and $\sim$1.2 mJy at 1 mm within the inner 40\% of the observed area, and $\sim$0.5 mJy at 3 mm and $\sim$4 mJy at 1 mm within the inner 90\% of the imaged area. The exact values can be found in Table \ref{tab:RMS}.

We then closely examine by eye parts of the image that contain extended structure and no compact emission, primarily \hii regions. We draw polygons\footnote{The polygon regions are available on GitHub with the rest of the reduction code: \url{https://github.com/nbudaiev/SgrB2_ALMA_continuum}} around these locations by hand so that any detections within them can be excluded. In a few cases, we found core-like sources within the \hii regions. We drew the polygons in a way that would include such sources in the final catalog.

Finally, we review each of the cataloged sources in both pointings and frequencies by hand. 
We rank every source as not-a-core (0), core candidate (1), or core (2). We inspect each source twice: first, zoomed-in closely to examine the shape and the background features and then using the whole image to get context on the source morphology based on location and proximity to other sources. 

For example, if a source with a score of 1 that resembles a filamentary structure is located close to an area with extended emission, such as an \hii region, we mark it as a false detection and adjust the score accordingly. If such a source is located in a cluster of other sources with a score of 2 and is not visibly a part of any extended structure, we change its score to 2.
We further investigate each source with a score of 1 by comparing their structure in the 1 mm and 3 mm data. If a potential source is part of a filamentary structure at 3 mm, but no local peak is observed at 1 mm, we score such a source as a 0. Such a filamentary structure at 3 mm without a strong counterpart at 1 mm is indicative of free-free emission.

All of the extended \hii regions are excluded manually using polygons as described above.
In addition, we cross-match our detections with \cite{DePree1998} and \cite{Meng2022} to remove known compact \hii regions from our core sample and label them as known \hii regions.
We find that VLA-detected \hii regions do not perfectly align with \hii region candidates in our data, especially in Sgr B2 M\refreport{: our sources are generally located at a greater declination.} 
\refreport{The most likely cause is the wrong positional coordinates of the phase calibrator J1744-3116 in VLA catalogs. When compared to other catalogs \citep[e.g.][]{Lanyi2010}, the location of J1744-3116 is offset by $\sim$0.29$''$ primarily in declination. The erroneous coordinates of the VLA phase calibrator has likely affected many of the previous observations of Sgr B2.} 
%Removed: We attribute this to poor absolute pointing accuracy of the VLA \citep{Mills2018}. 
When it is not possible to establish a confident one-to-one match between catalogs, we label the source as an \hii region candidate. 
Some of the previously known \hii regions appear to have multiple peaks within their areas. We label these ``fragmented" sources as \hii region candidates. 

To further identify \hii regions based on their spectral properties, we create a 1 mm to 3 mm spectral index map of the central, most dense parts of Sgr B2 N and M. We find that the central region of the Sgr B2 M, being more evolved compared to Sgr B2 N, has spectral indexes close to 0, which is typical for optically thin \hii regions. Considering the high density of \hii regions in the same area described in \cite{Schmiedeke2016} we mark all previously uncategorized sources in the close vicinity \refreport{of} \hii region candidates and give them a score of 0. Only a few sources are removed in Sgr B2 N with this method due to the high error values of the spectral index map in the region.
We label the sources \refreport{that} morphologically \refreport{resemble} \hii regions (e.g. a source with a dim center but bright edges) as \hii region candidates. Finally, we mark sources that are extremely bright at 3 mm as \hii region candidates. 
Every \hii region candidate that is not a previously known \hii region, cannot be assigned to a previo\refreport{u}sly known \hii region
% removed: accounting for VLA's poor absolute pointing accuracy
, and is not a part of a ``fragmented" known \hii region is marked as a new \hii region detection.
In total, our catalog contains 17 known hyper-compact and ultra-compact \hii regions, 16 \hii region candidates, and 8 new hyper-compact \hii regions. \refreport{5 of the new \hii regions are detected due to increase in resolution compared to VLA-based observations. The other 3 new \hii regions are located within Sgr B2(N)-SMA2 hot core \citep{Qin2011}.
}

Having two criteria for core identification, strict and lax, allows for separate ``robust" and ``complete" catalogs to co-exist. We note that the difference between a score of 1 and 2 is subjective. 
After performing the analysis described in the following section, we do not find any significant differences when comparing the results between the two catalogs. Thus, we report and use only the ``complete" catalog in the rest of this work. Our catalog contains 371 cores at 3 mm and \refreport{218} cores at 1 mm, of which \refreport{179} have a 3 mm counterpart. Thus, the total number of individual cores identified at at least one wavelength is \refreport{410}. 

\begin{figure}[th]
    \centering
    \includegraphics[width=\columnwidth]{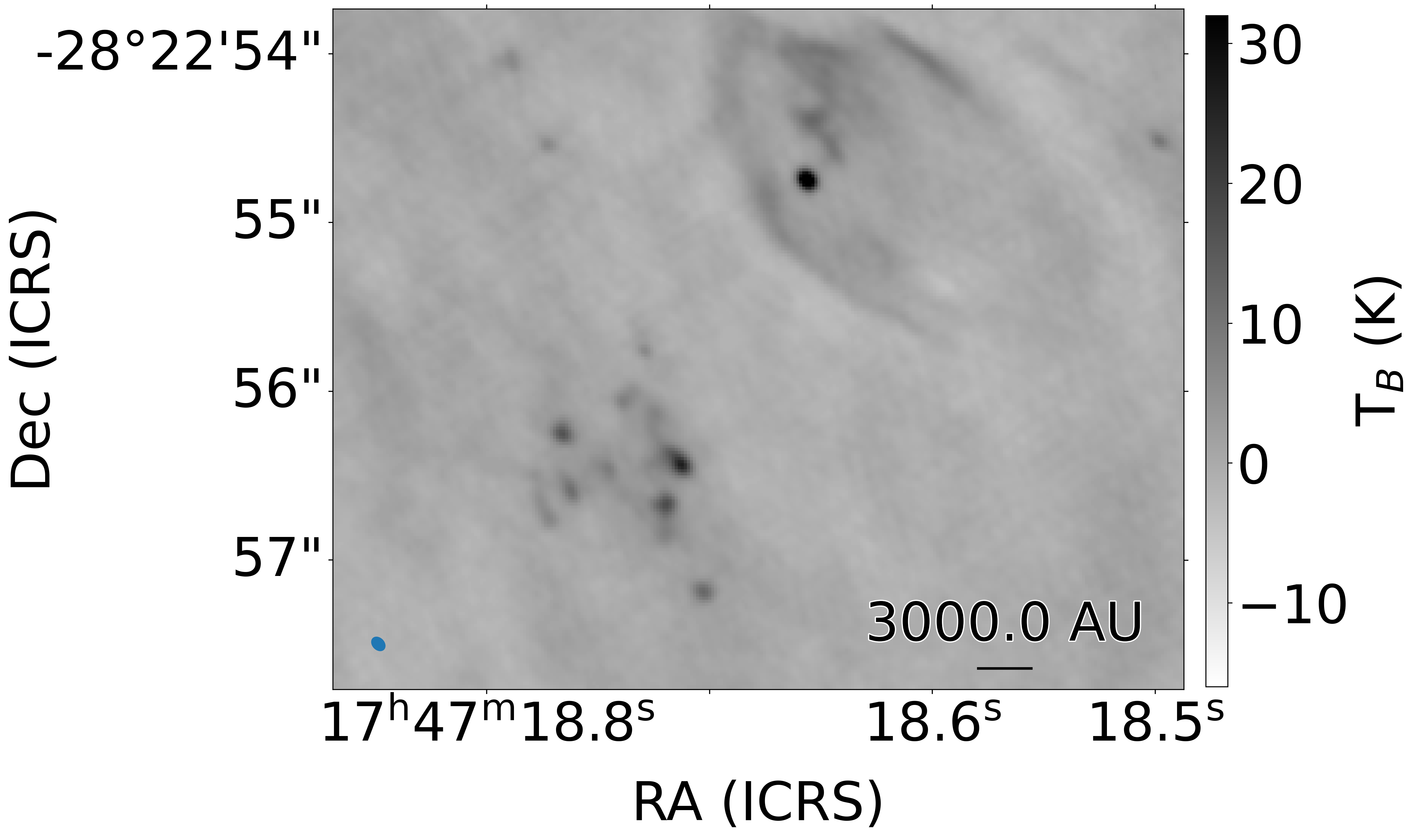}
    \caption{A zoom in on a region north-east from Sgr B2 M at 3 mm. It features \hii region Sgr B2 Y from \cite{Schmiedeke2016} on the top-right, a core that is within the \hii region's projected area on the sky, and a highly-fragmented source on the bottom-left. This highly-fragmented group of ten detected cores was previously thought to be one massive core. }
    \label{fig:zoomin}
\end{figure}

The described approach has some caveats.
Even after extensive self-calibration and cleaning, the data still include local artifacts such as negative bowls around bright sources and wave-like patterns on different scales. 
The high source density and the presence of extended emission (e.g. Figure \ref{fig:zoomin}), complicates the estimation of the local noise for each source.

Excluding detections within polygons that were drawn by eye can potentially exclude true positives. While it is unlikely for cores to be present inside \hii regions, they might appear to be so due to projection effects. Less than a few percent of the total pointing area is excluded by the polygons. Potentially missed sources should not affect the conclusions of this paper as there would be no biases for the excluded sources.

\subsection{Photometry}
\begin{figure}[th]
    \centering
    \includegraphics[width=\columnwidth]{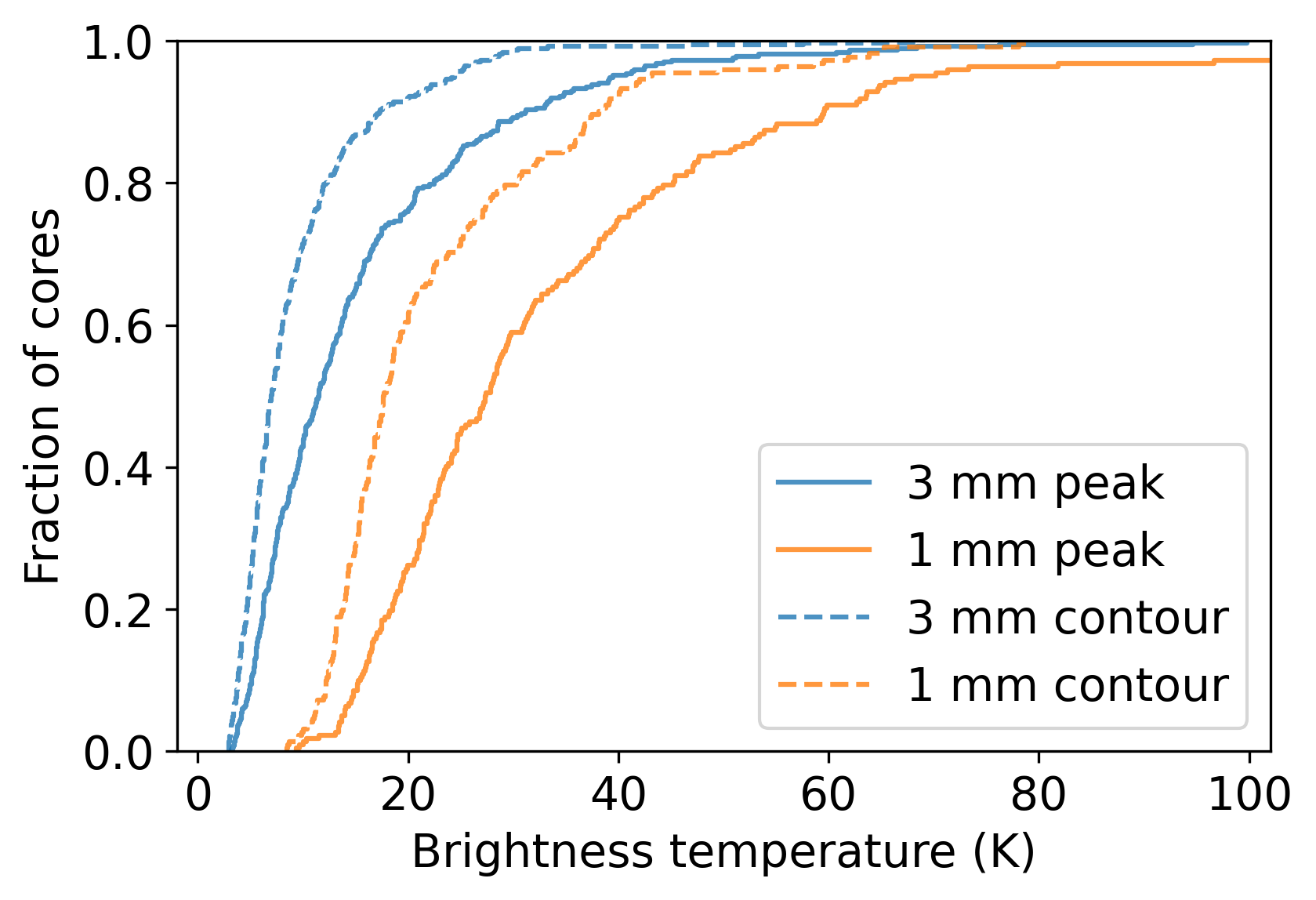}
    \caption{Brightness temperature of the cataloged cores. The brightness temperature is serves as a lower limit  of the kinetic temperature for blackbody sources.}
    \label{fig:brightness_T}
\end{figure}

We use the integrated flux within the dendrogram structure contour and the brightest pixel within the contour as the primary observables. We present the two measurements of source flux in the catalog, a sample of which is shown in Table \ref{tab:catalog}. The full catalog is available in machine-readable format. The source brightness temperatures are shown in Figure \ref{fig:brightness_T} as a cumulative distribution function (CDF).
The total source flux is the sum of the pixel values in Jy in the dendrogram-defined contour referred to as a ``leaf". The peak is the value of the brightest pixel in Jy/beam. These numbers are expected to be the same for a point source and the peak-derived value should be smaller for a resolved source. Dendrogram-based flux underestimates the flux of the very faint sources because the lowest possible contour is set at a given threshold level, in our case 4$\sigma$.

During testing, we attempted to perform 2D Gaussian photometry using the \texttt{gaussfit\_catalog} python package.
We inspected each fit by hand and found that fewer than 10\% of the sources had a satisfactory fit for the central object. We found that many of the fits were affected by the presence of non-uniform extended envelopes around the centrally peaked sources, like in Figure \ref{fig:zoomin}. Furthermore, high source density and imaging artifacts also contributed to some of the failed fits. Thus, we concluded that our sources are not well-represented by simple 2D Gaussians and decided to use dendrogram contours to define sources.

\subsubsection{Photometry comparison} \label{sec:photcomp}
As a part of Cycle \refreport{2} ALMA project 2013.1.00269.S, \citet[][hereafter G18]{Ginsburg2018} obtained Band 3 observations of the extended Sgr B2 cloud at 0.5" resolution.

\begin{figure}[th]
    \centering
    \includegraphics[width=\columnwidth]{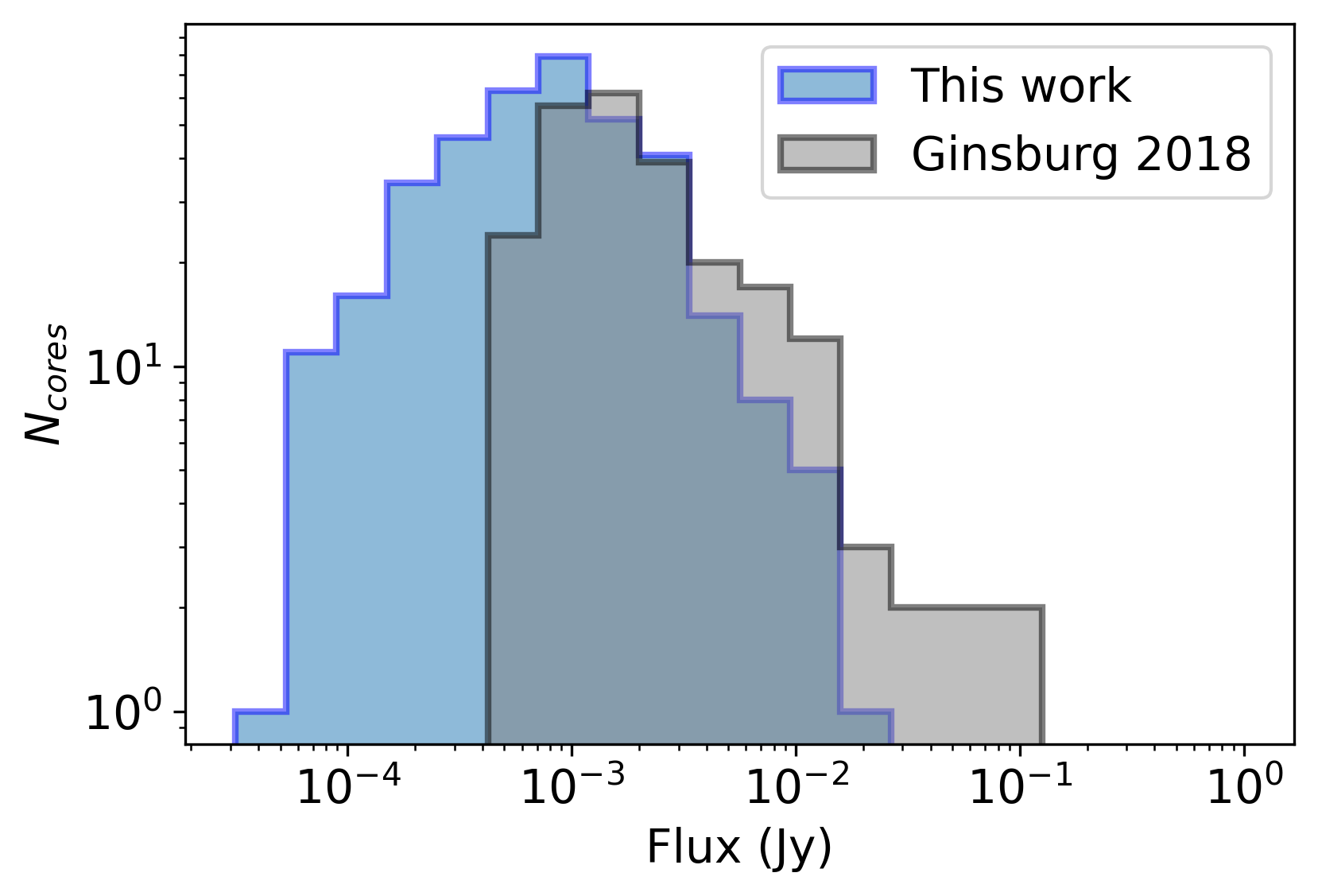}
    
    \caption{A comparison of our dendrogram contour-based, 3 mm source flux and the full 15$\times$15 pc G18 catalog, with known \hii regions removed from both data sets. The fluxes from G18 are \refreport{multiplied} by 0.92 to account for the difference in spectral setup assuming the sources have a spectral index of two. 
    The general shape appears to be similar, but with the higher-resolution data, which have twice the sensitivity, we are able to detect fainter sources. The turnover point is at about 1 mJy, which is just below the turnover point in G18 catalog.}
    \label{fig:fluxes}
\end{figure}

We investigate the impact of the differences between ours and G18 observations.
Compared to G18, the catalog derived in this work has a factor of 10 better spatial resolution and is a factor of two more sensitive.
The distribution of source fluxes is shown in Figure \ref{fig:fluxes}.
The data from G18 were centered at 96.36 GHz with \refreport{each of the four spectral windows having a} bandwidth of 1.875 GHz, while our observations are centered at 92.45 GHz with \refreport{four spectral windows of the same bandwidth}. 
Since the central frequencies of the data are not the same, we perform a frequency correction.
Assuming that the majority of the flux within the G18 beam is coming from the optically thick part of the core or an \hii region and it has an average spectral index of two, the frequency correction factor is $\sim$0.92. For a positive spectral index, the flux measured from our data within the defined aperture is expected to be \refreport{dimmer}. Depending on the source, the correction factor can range from 1.05 at spectral index of -1 to 0.87 at spectral index of 3.5 which would correspond to about 15\% flux correction error.
As shown in Section \ref{sec:spectral_indexes}, many of the cataloged sources have spectral indexes between 2 and 3.

\begin{figure}[ht]
    \centering
    \includegraphics[width=\columnwidth]{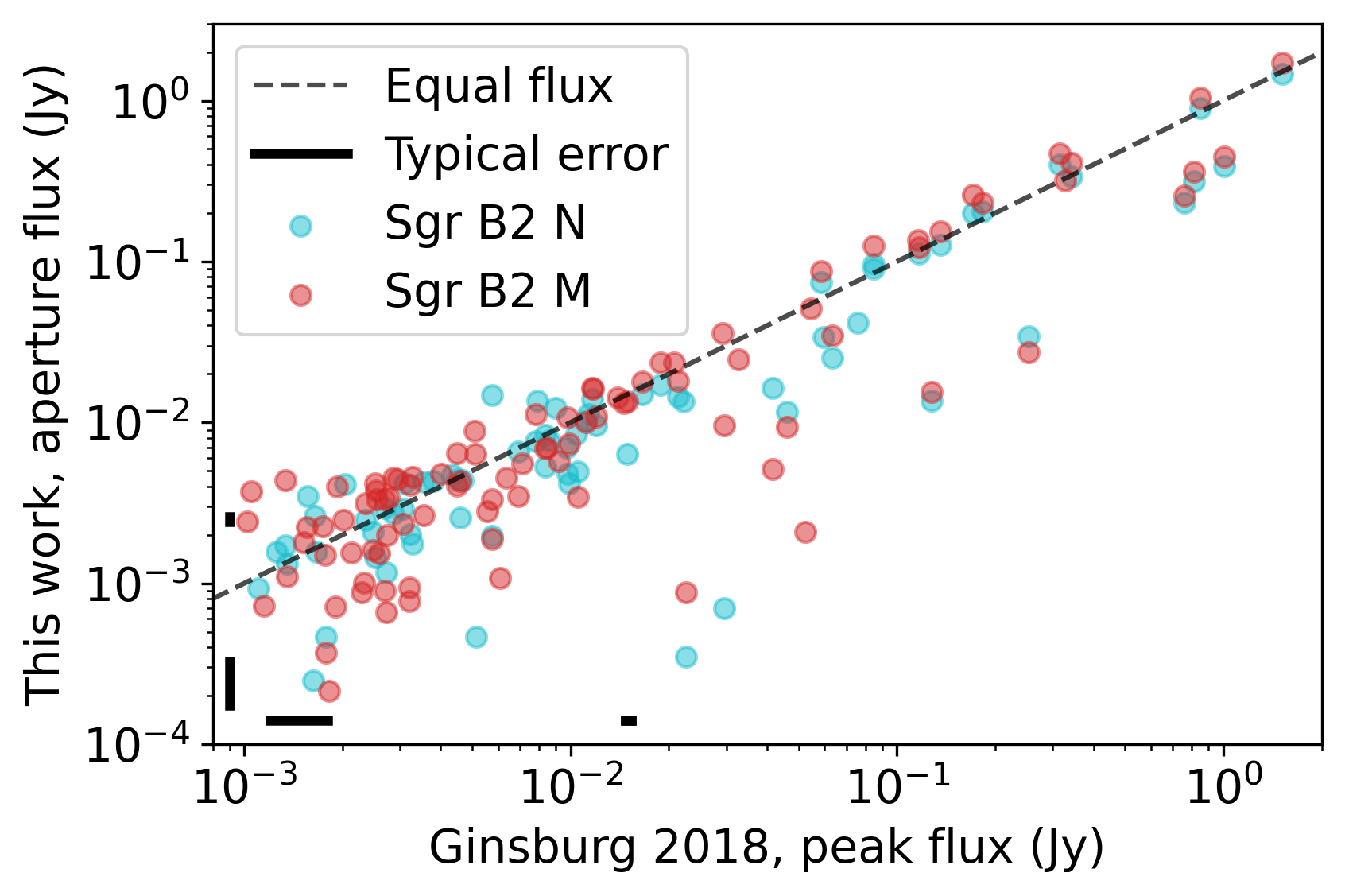}
    \caption{Aperture photometry comparison of our data and \refreport{G18} data. The typical flux errors are shown as solid black lines at different fluxes on the log scale. Some scatter is present that can be explained by the difference in resolution, frequency, and the presence of imaging artifacts. \refreport{The sources well below one-to-one flux ratio are due to negative bowls in the high-resolution data.} The data points from Sgr B2 N and M that are almost identical are caused by the overlap in the two pointings' field of view.}
    \label{photometry}
\end{figure}

We compare our and G18 images by performing aperture photometry on G18 data.
We apply an elliptical aperture equivalent to 1-$\sigma$ of the G18 Gaussian beam (0.230" $\times$ 0.197") on the locations of G18 detections on our data and compare it to the ``peak" (brightest pixel) flux from G18.
However, an elliptical aperture does not capture all the flux counted in the brightest pixel of an unresolved source. While the brightest pixel is representative of the sum of the Gaussian distribution of the signal, an elliptical aperture cuts off the edges of the distribution. We calculate the aperture correction factor as the fraction of the Gaussian distribution that is cut off after using a 1-$\sigma$-sized aperture to be 1.465.
Finally, we convert the summed flux from Jy/beam to Jy.
We compare this quantity with the brightest pixel of G18 sources. The ratio of flux at the location of each G18 source is shown in Figure \ref{photometry}. 
There is a slight trend of higher-resolution sources having less flux. A smaller fraction of sources are above the line. Low-flux sources exhibit larger scatter, which is consistent with measured noise.
The somewhat large scatter can be partially explained by the presence of imaging artifacts in the images, primarily negative bowls. \refreport{All data points with flux ratio of over 10 are due to negative bowls in the high-resolution data.}.

\subsection{Source sizes} \label{sec:source_sizes}
\begin{figure}[ht]
    \centering
    \includegraphics[width=\columnwidth]{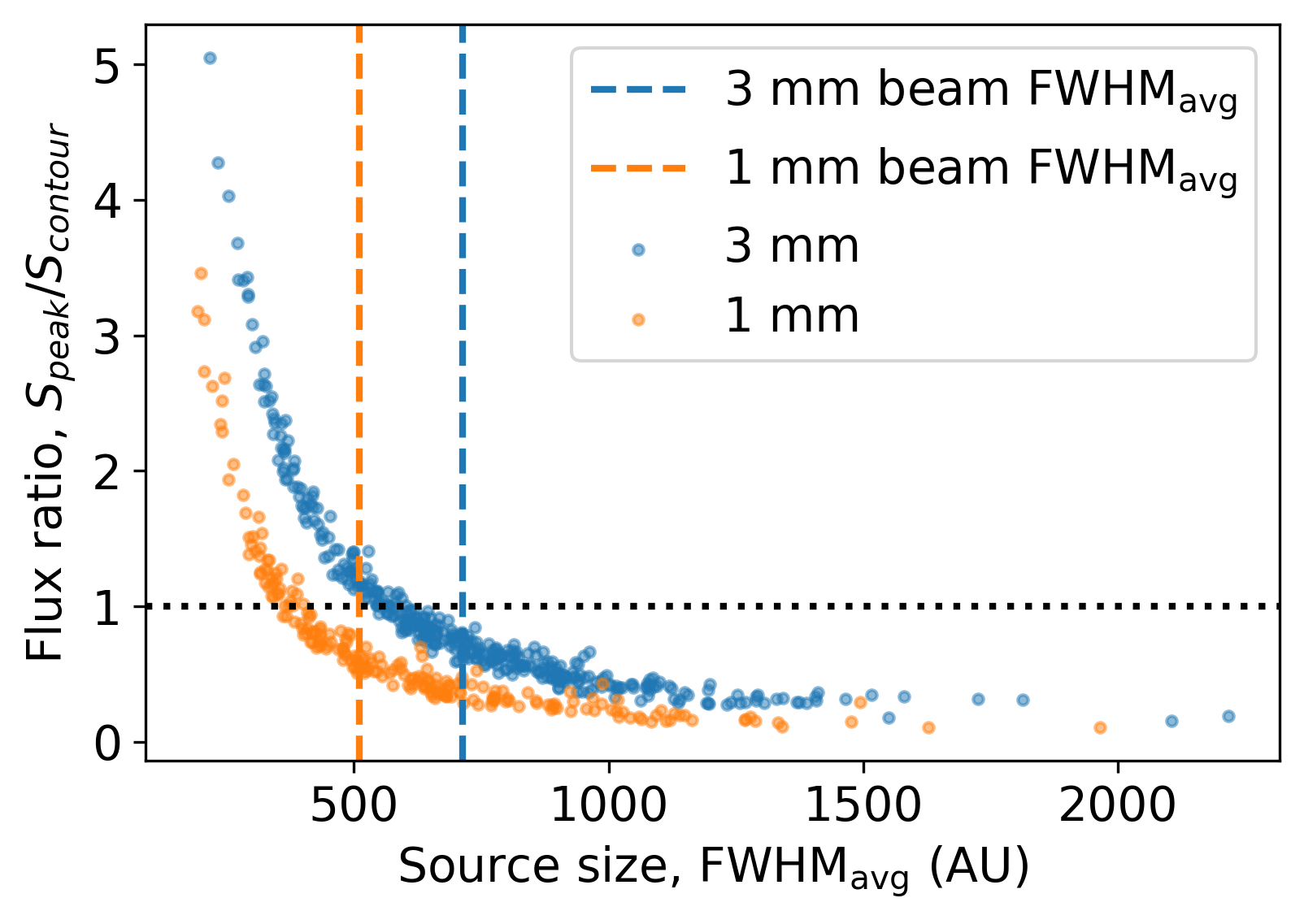}
    \caption{The ratio of the brightest pixel and the total source flux within dendrogram leaf structure is plotted against the average source FWHM as recorded by \texttt{astrodendro}. 
    The flux ratios below one indicate likely resolved sources, which represent a majority in our sample.}
    \label{source_size}
\end{figure}

For each source we recorded the total flux within the dendrogram object contour and the brightest pixel value within the contour. If a source is not resolved, then the brightest pixel should contain all the flux from the source. If a source is resolved, the brightest pixel will contain only part of the flux. Thus, we use a ratio of brightest pixel flux and summed flux within dendrogram structure to serve as a metric of whether a source is resolved or not. As shown in Figure \ref{source_size}, we find that $20\%$ of 1 mm and $30\%$ of 3 mm sources have effective areas smaller than the beam area and peak flux larger than the dendrogram contour flux, which means that the sources are unresolved.
On the other hand, only a handful of sources are more than two resolution elements across -- 23 sources at 1 mm and 8 sources at 3 mm are well resolved.

The sources with $S_{peak} / S_{contour}$ $<$ 1 and dendrogram contour area larger than the beam size are marginally resolved. There are 45$\%$ of such sources at each wavelength.

\subsection{Spectral indexes} \label{sec:spectral_indexes}
\begin{figure*}[ht]
        \fig{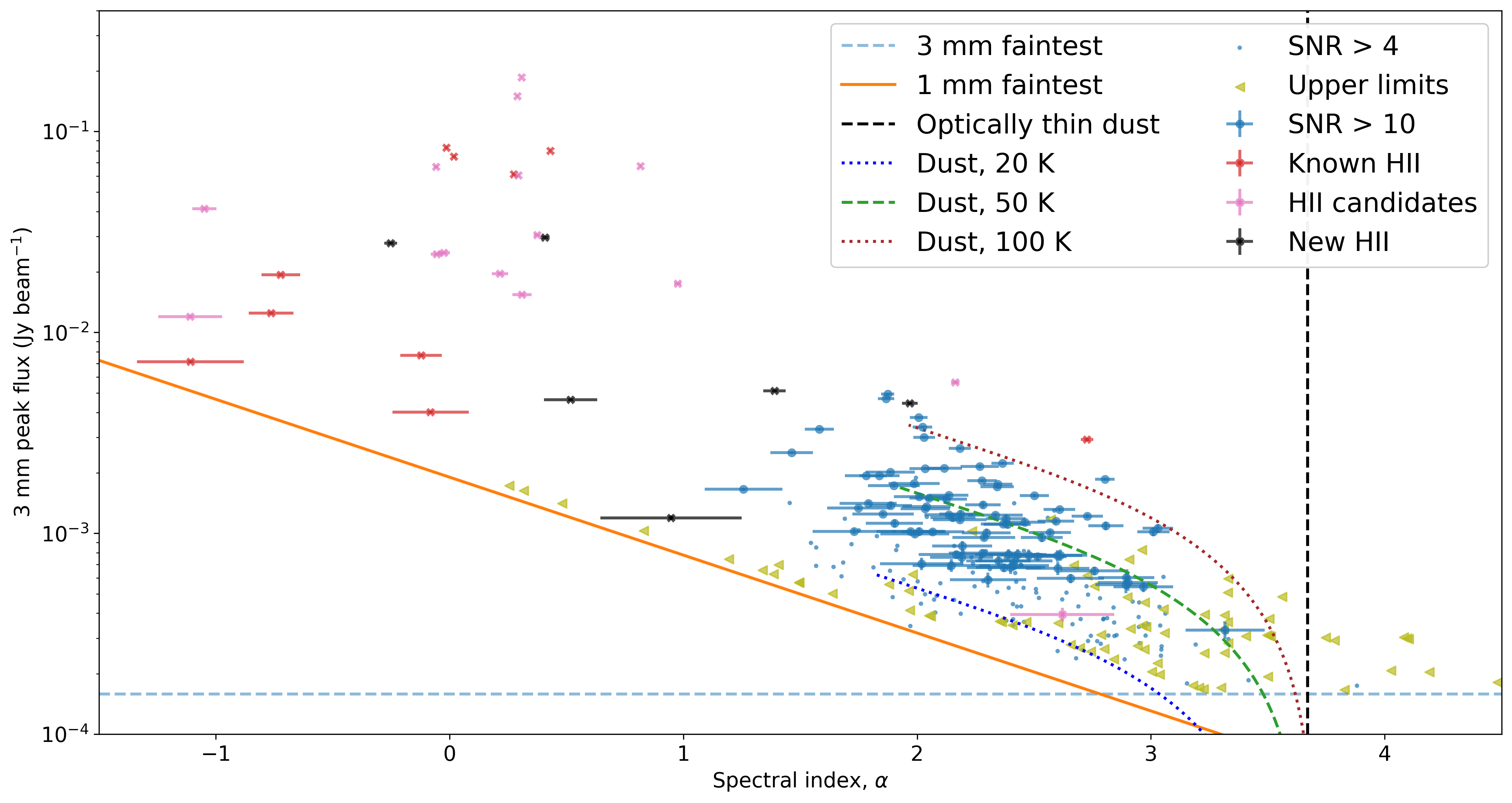}{0.88\textwidth}{(a)}
        
        \fig{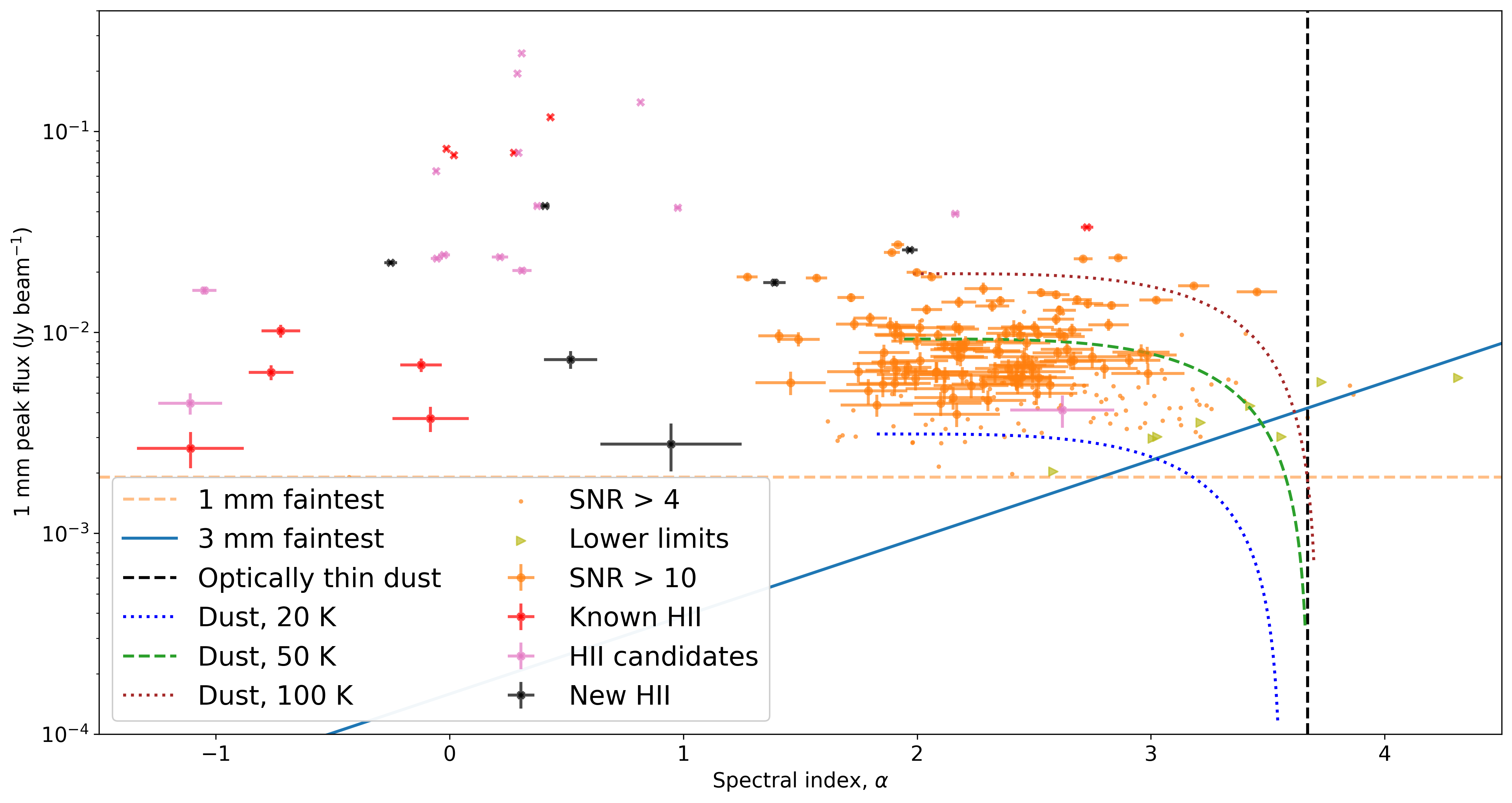}{0.88\textwidth}{(b)}
    \caption{Source flux of 3 mm (a) and 1 mm (b) ALMA detections plotted against the calculated spectral index using the corresponding location at the other wavelength after matching beam sizes. We only calculate the errors of sources with signal-to-noise above 10, but all sources are shown for completeness. Each line represents a physical or observational limit on possible values. The 1 mm and 3 mm flux limits are based on the faintest detected source in the corresponding band. The dashed-green and dotted lines show the maximum flux at different temperatures of a source at different spectral indexes assuming the source is beam-sized \refreport{following the modified blackbody Equation \ref{eqn:mbbe}}. 
    \refreport{We assume $\beta_{\text{dust}} = 1.75$ which puts optically thin T = 50 K dust at ${\alpha = 3.66}$}. The detected sources fill the observable range completely. Follow up observations with higher sensitivities may reveal the missing dusty, optically thin sources. The olive triangles show the upper and lower limits on the spectral indexes for sources that have at least $4\sigma$ signal only in one band. The flux in the other band was taken as $4\sigma$ at that location in the image. The flux error bars are generally much smaller than the marker size and are only visible for low-flux sources.}
    \label{fig:SI}
\end{figure*}

In order to measure the spectral indexes, we need to convolve the data to a common beam.
The common beam between the two bands is equivalent to the Band 3 beam. Thus, we convolve the Band 6 data to the Band 3 beam, accounting for the beam area change as our data are in units of Jy/beam.
To calculate the spectral index, we take the brightest pixel values for detections at 3 mm and take the corresponding positions on the sky at 1 mm. 
The brightest pixel contains information about at least the inner, most dense part of the source in its 2D projection.
The mean offset between the locations of the brightest pixels at 3 mm and 1 mm is less than a quarter of the beam, which is equivalent to $\sim$1 pixel and is never more than one common beam apart. We find that using the 1 mm data's brightest pixel locations produces very similar individual measurements, much less than the calculated errors.
We check for systematic errors that could be caused by this approach and find no trends in the measurements of the two distribution\refreport{s}.
In addition to calculating spectral indexes for sources present in both bands, we are also able to place upper and lower limits for sources detected only at 3 mm and 1 mm respectively. We show the distribution of spectral indexes with the corresponding source fluxes in Figure \ref{fig:SI}, indicating physical and observational constraints.

% deleted: Since the flux measurements in two bands are uncorrelated distributions, 
We calculate the spectral index errors only for sources with SNR over 10 in both bands. 
For sources with SNR below 10 in either band, the errorbars are not necessarily representative of the uncertainty because the error distribution function is no longer Gaussian (the distribution function for the ratio of two Gaussians with mean zero is a Cauchy distribution, for which the standard deviation and mean are undefined). We therefore show data points without error bars for these sources in Figure \ref{fig:SI}. 
All sources in our catalog have the SNR $>$ 4.
%removed
%Convolving Band 6 data to the common beam results in a factor of two decrease in SNR for compact sources, which leaves only a few sources with SNR above 10 and significantly decreases the number of sources with SNR $>$ 4. 
%Assuming that the radial brightness profile of the sources is the same, convolving data to a larger beam would increase the source flux measurements by the same factor. The resulting flux distribution of sources with the new, lower SNRs would be scaled, but have a similar shape.
%Thus, we use the original SNR, before data convolution, to perform the 10$\sigma$ and 4$\sigma$ cuts. We then use the convolved data to calculate individual source errors.
Convolving Band 6 data to the common beam decreases the SNR for compact sources. To maximize the number of included Band 6 sources, we perform the 10$\sigma$ and 4$\sigma$ cuts based on the original SNR. We then use convolved data to calculate individual source uncertainties.

We investigated whether we could use the alpha-maps produced by the \texttt{tclean} algorithm to estimate the in-band spectral indexes. Fewer than a dozen sources had errors below $50\%$. Thus, our data is not sensitive enough to produce in-band spectral indexes.

\begin{figure}[th]
    \centering
    \includegraphics[width=\columnwidth]{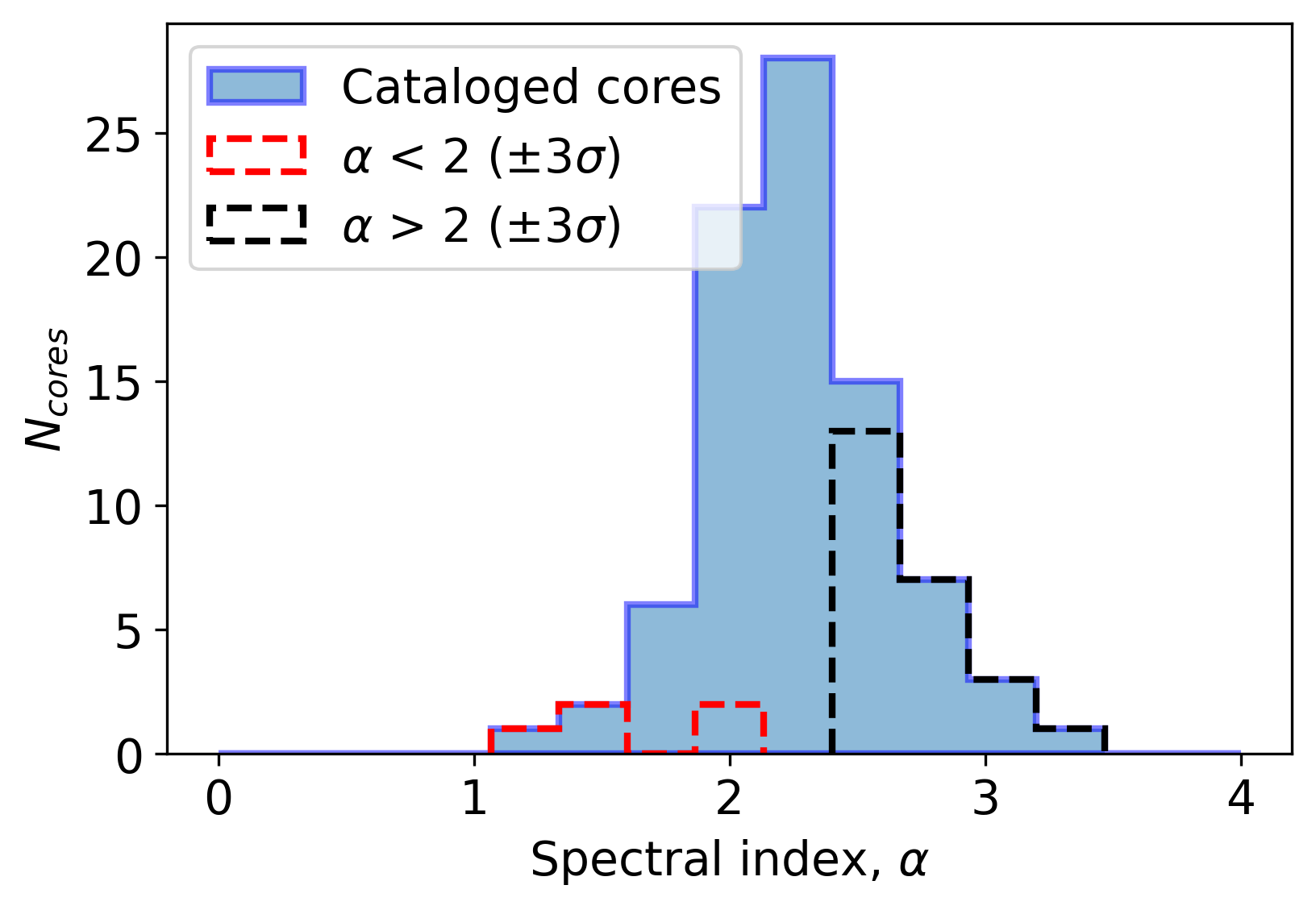}
    \caption{Spectral indexes of the sources in our catalog. We cross-matched 92.45 GHz and 225.78 GHz detections and calculated the spectral indexes as described in Section \ref{sec:spectral_indexes}. \refreport{Here, }we exclude any source that has the a continuum signal-to-noise below 10 in either band. The red dotted histogram shows the sources that have been cataloged as ``cores" that have statistically significant ($\pm3\sigma$) spectral index below 2. The black dotted histogram shows the fraction of sources that are not completely optically thick and have the calculated spectral index above 2.}
    \label{fig:SI_hist}
\end{figure}

We explore the potentially misclassified sources in our catalog.
The ``core" classification was done based on the morphology, proximity to other sources, relative brightness, and surrounding emission. The spectral index values can then be used to identify potential catalog contaminants such as \hii regions. As shown in Figure \ref{fig:SI_hist}, there are five sources that have statistically significant spectral indexes below 2, which are probably misclassified \hii regions. 
However, at least three of these sources are located in the negative bowls in Band 6 data. Since our cataloged fluxes are not background subtracted, being located in the negative bowl causes a lower 1 mm flux and underestimates the spectral index. 
Similarly, being located in a negative bowl at 3 mm artificially increases the spectral index. A known \hii region Sgr B2 F10.30 is the most prominent example of a source in a 3 mm negative bowl, having $\alpha$ = 2.7. Thus, some of the detections near negative bowls might have been misclassified; as shown in Figure \ref{fig:SI}, there are some \hii region candidates that have $\alpha > $ 2 and cores with $\alpha < 2$.
Fewer than 1\% of sources are located \refreport{in a of negative bowl}.
% in a vicinity of negative bowls -> in a negative bowl

We also note that just above 60\% of our 3 mm detections fall within the field-of-view of 1 mm pointings. We extrapolate the derived spectral-index-based conclusions to the rest of the 3 mm sources.
While the distance from the central regions of the clusters might have some impact on core temperature and age, it does not impact the conclusions of this paper.

\subsection{Source masses} \label{sec:source_masses}
Using a number of assumptions about the properties of a core, including that the dust is optically thin, we can derive the mass of such a core. 
We start from a modified blackbody equation:
\begin{equation}
    B^{'}_{\nu} = \frac{2 h\nu^3}{c^2} \left(e^{h \nu / k_B T}-1\right)^{-1}\left(1-e^{-\kappa_\nu\Sigma}\right),
\label{eqn:mbbe}
\end{equation}

where $B^{'}_{\nu}$ is the spectral brightness of a modified blackbody, $h$ is the Planck constant, $\nu$ is the frequency, $c$ is the speed of light, $k_B$ is the Boltzmann constant, \refreport{$T$ is the kinetic temperature of the source,} $\kappa_\nu$ is the dust opacity index, and $\Sigma$ is the surface density.
Given an area $A$, the mass $M$ is
\begin{equation}
    M = \Sigma A = \Sigma \pi r^2,
\end{equation}
where $r$ is the source radius.
%\begin{equation}
%    S_\nu = B_\nu \frac{\pi r^2}{d^2}
%\end{equation}
Converting the spectral brightness $B^{'}_{\nu}$ to the observed spectral flux density $S_\nu$ and applying the optically thin dust assumption ($\tau = \kappa_\nu \Sigma << 1$):
\begin{equation}
    S_\nu = \frac{2 h\nu^3}{c^2}\left(e^{h \nu / k_B T}-1\right)^{-1}\frac{\kappa_\nu M}{\pi r^2} \frac{\pi r^2}{d^2},
\end{equation}
where $S_\nu$ is the observed flux density, $d$ is the distance to Sgr B2, and $M$ is the source mass.
Simplifying using the Planck's law and solving for mass yields: 
\begin{equation}
    M = \frac{S_\nu d^2}{B_\nu\kappa_\nu},
\end{equation}
where $B_\nu$ is the spectral brightness of a blackbody.
Adopting the Rayleigh-Jeans approximation and expressing in terms of brightness temperature:
\begin{equation} \label{eqn:mass}
    %M = \frac{S_\nu d^2 c^2}{2 \kappa_\nu \nu^2 k_B T}
    M = \frac{T_B d^2}{\kappa_\nu T_{KE}},
\end{equation}
where $T_B$ is the brightness temperature and $T_{KE}$ is the kinetic temperature of the source. 
Extrapolating the \cite{Ossenkopf1994} opacity models for dust grains with thin ice mantles, gas density of $10^6\ \mathrm{cm}^{-3}$, and age of $10^5$ years as a power law, assuming $\beta_{\text{dust}} = 1.75$, we obtain $\kappa_{92 GHz} = 0.0017\ \frac{cm^2}{g}$ and $\kappa_{226 GHz} = 0.0083\ \frac{cm^2}{g}$.
Further assuming a uniform core temperature $T_{KE} = 50$ K and a distance\footnote{Using the most recently measured distance of 8.127 kpc \citep{GRAVITYCollaboration2018} would result in 7\% lower masses.} of 8.4 kpc we \refreport{can calculate source masses}.

\begin{figure}[th]
    \centering
    \includegraphics[width=\columnwidth]{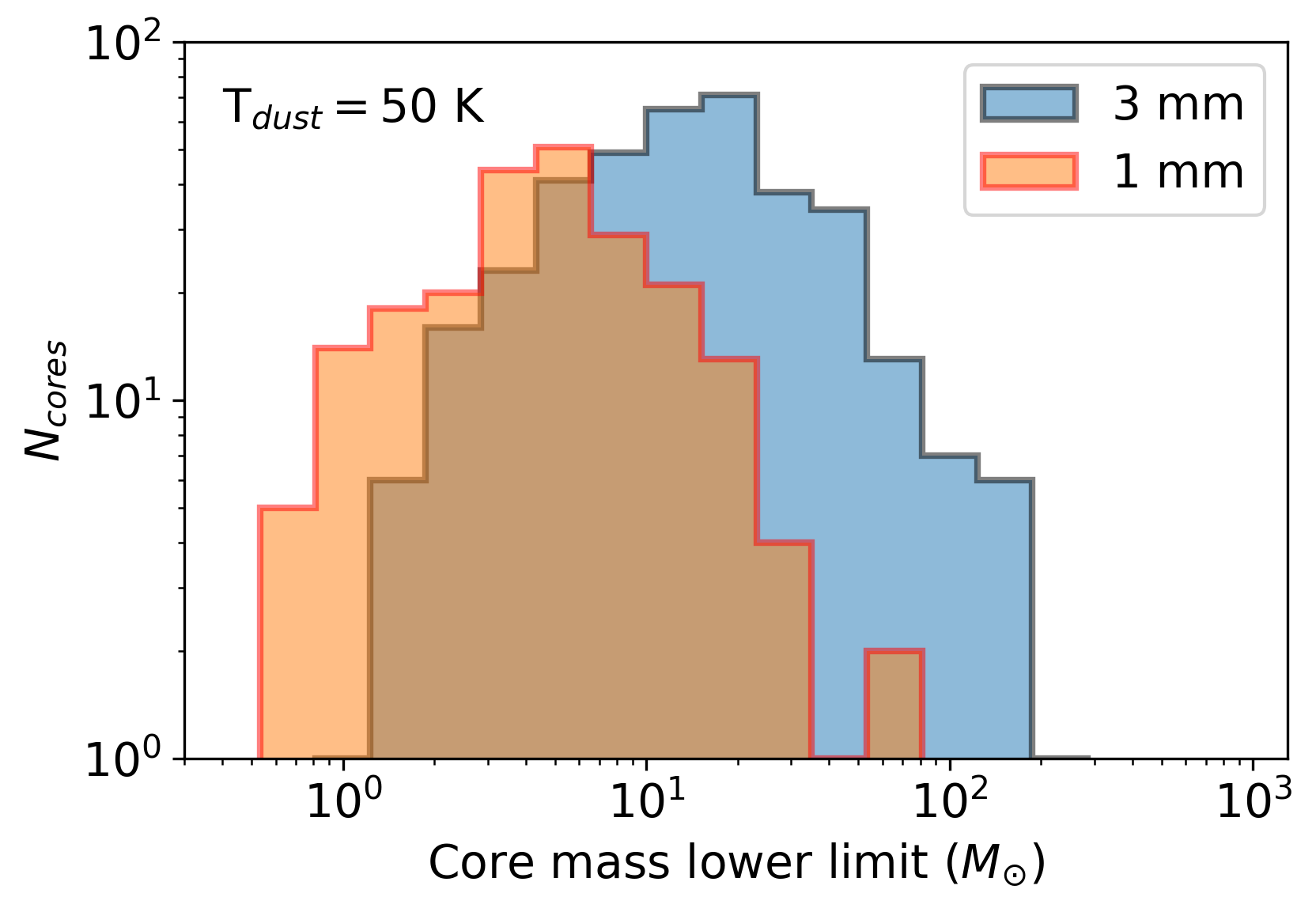}
    \caption{The inferred lower mass limit of the sources using the optically thin dust assumption for 1 mm and 3 mm data. 1 mm data has a higher mass sensitivity that is caused by the dust becoming more optically thick at higher frequencies and thus lower mass sources can be detected. We assume a uniform dust temperature of 50 K and a gas-to-dust ratio of 100. The turnover point at both wavelengths originates from the completeness limit over the whole imaged area.}
    \label{fig:masses}
\end{figure}

\refreport{Here, we utilize the total dendrogram ``leaf" flux instead of the source's brightest pixel value to account for the sources that are partially resolved. 
We then incorporate an aperture correction factor for the source flux. The \texttt{astrodendro} algorithm excludes any emission that is below the specified \texttt{min\_value}, in our case 4$\sigma$. Since the majority of our sources are unresolved or marginally resolved, we can assume 2D Gaussian flux distribution and account for the cut-off flux. 
For very crowded regions, \texttt{astrodendro} leaves might not extend to the minimum flux value. Thus, even with this correction some masses might be underestimated.}
In addition, since the cores are at least partially optically thick, we can obtain only a lower limit on their dust masses.
Other variables being constant, an optically thick source contains more mass than an optically thin source. 
\refreport{The calculated core masses are presented in Figure \ref{fig:masses}.}

Both the individual mass measurements and mass distributions based on the 1 mm and 3 mm observed flux do not match. We attribute this to the erroneous assumption of optically thin dust used for this calculation. Such an assumption underestimates the masses at 1 mm more than at 3 mm as dust has longer mean free path at longer wavelengths and thus flux is emitted from more of the material before reaching optically thick regime. 
Thus, we use the masses based on the 3 mm data as the lower limits in the rest of the paper. The core mass distributions have the turnover point \refreport{around} our completeness threshold over the whole imaged area \refreport{at 6$\sigma$ (13.5 \msun)}.
\refreport{Using \texttt{plfit}, a Python implementation of the general power law fitting algorithm \citep{Clauset2009}, we fit a power law to the masses above the completeness limit. The resulting fit of 191 sources is $\alpha = 2.4\pm 0.1$, which is consistent with the Kroupa IMF $\alpha = 2.35$. If we increase the fitting threshold to 8$\sigma$ (18 \msun), the 144 sources produce a slightly steeper fit $\alpha = 2.6\pm 0.1$.}

Using 50 K uniform core temperature gives an understanding of possible mass ranges and shows the general shape of the core mass distribution.
As shown by \cite{Bonfand2019}, some cores can have average temperatures close to or above 150 K, usually referred to as hot cores. We detect sources at the locations of at least several such hot cores. Such sources are most likely the hot cores' central disks as our largest recoverable scale of $\sim$7500 AU is on the lower side of the typical hot core sizes \refreport{in Sgr B2} (Jeff et al., in prep).
At such high temperatures, the ice mantles surrounding the dust will start to sublimate. Such physical changes would decrease the emissivity by a factor of a few. We also do not expect a significant change to the gas-to-dust ratio based on freeze-out studies \citep{Caselli2022}.

\section{Discussion} \label{sec:disc}
\subsection{Nature of the observed sources} \label{sec:nature}

\subsubsection{Protostellar cores: Stage 0/I YSOs} \label{sec:YSO_I}
Our ``core" catalog consists primarily of compact dusty sources with sizes between 200 AU and 1000 AU.
Based on basic modeling, we show in this section that our cores are most consistent with an average temperature $\sim$50 K.
Such objects are, given the lower limit on their densities, unlikely to be in pressure equilibrium. We therefore identify these as rotationally-supported disks consistent with Stage 0/I YSOs\footnote{We extrapolate the original luminosity-accretion-based definition of stages for low-mass sources to their high-mass analogues from \cite{Robitaille2006}.}. We show the arguments in favor of this conclusion below.

In contrast to many studies of cores in nearby star-forming regions that assume a constant disk temperature of 20 K to make consistent comparisons between the works \citep[e.g.,][]{Eisner2018, Otter2021} we show that the core temperatures in Sgr B2 are more consistent with T = 50 K. 
Average dust temperatures based on \textit{Herschel} line-of-sight observations \citep{Etxaluze2013} and theoretical models of external dust heating due to stars \citep{Schmiedeke2016} both produce $\sim$20 K average temperature in the moderate-density dust ($n\lesssim10^5\ \mathrm{cm}^{-3}$) in Sgr B2 region. However, the denser regions have gas temperatures that are substantially higher \citep[e.g.,][]{Ginsburg2016, Krieger2017}. The modeled dust temperatures towards the centers of Sgr B2 N and M can go as high as 350 K and 600 K respectively \citep{Schmiedeke2016}. Since we are observing cores, which are local overdensities, we expect temperatures higher than 20 K.

\begin{figure}[th]
    \centering
    \fig{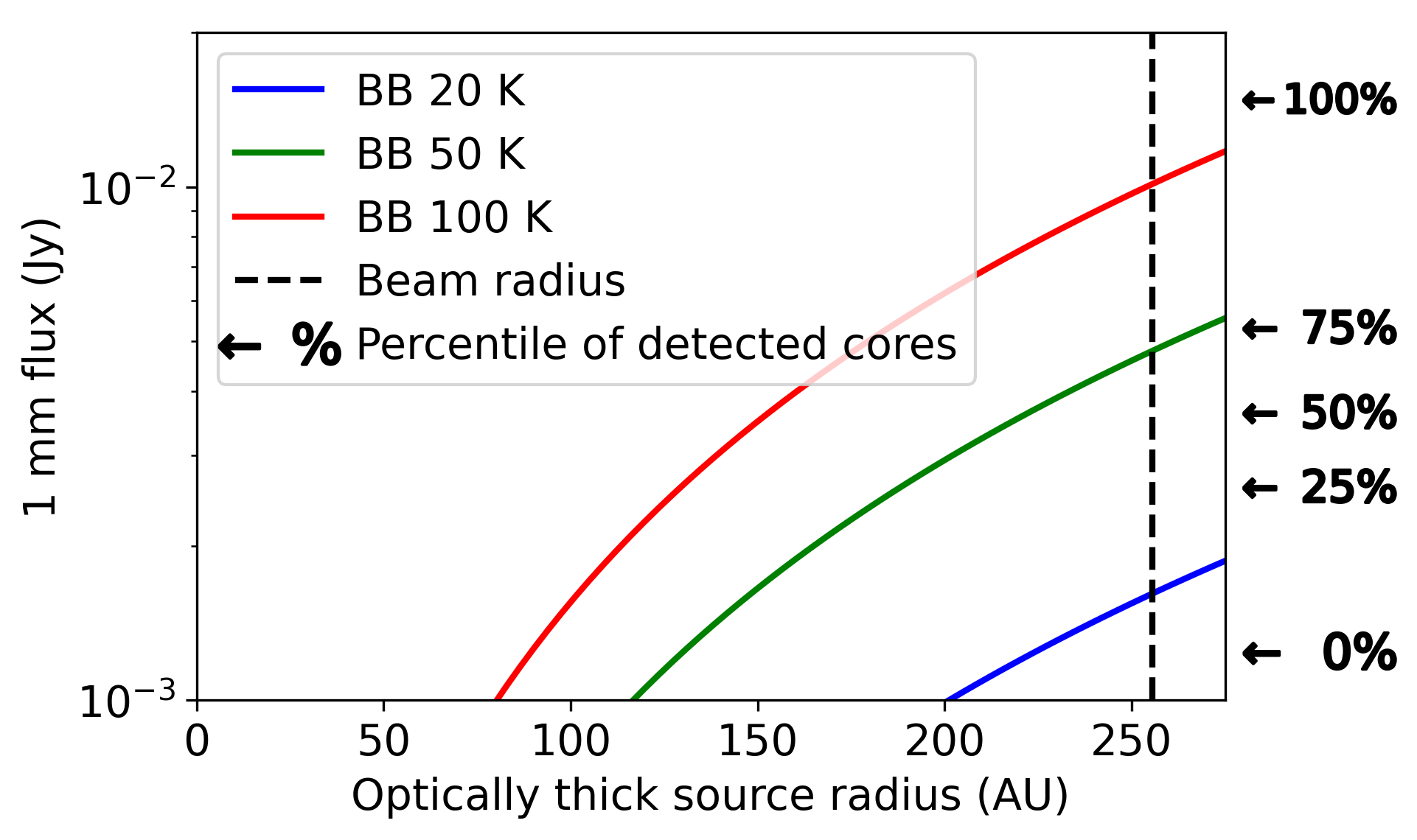}{\columnwidth}{(a)}
    
    \fig{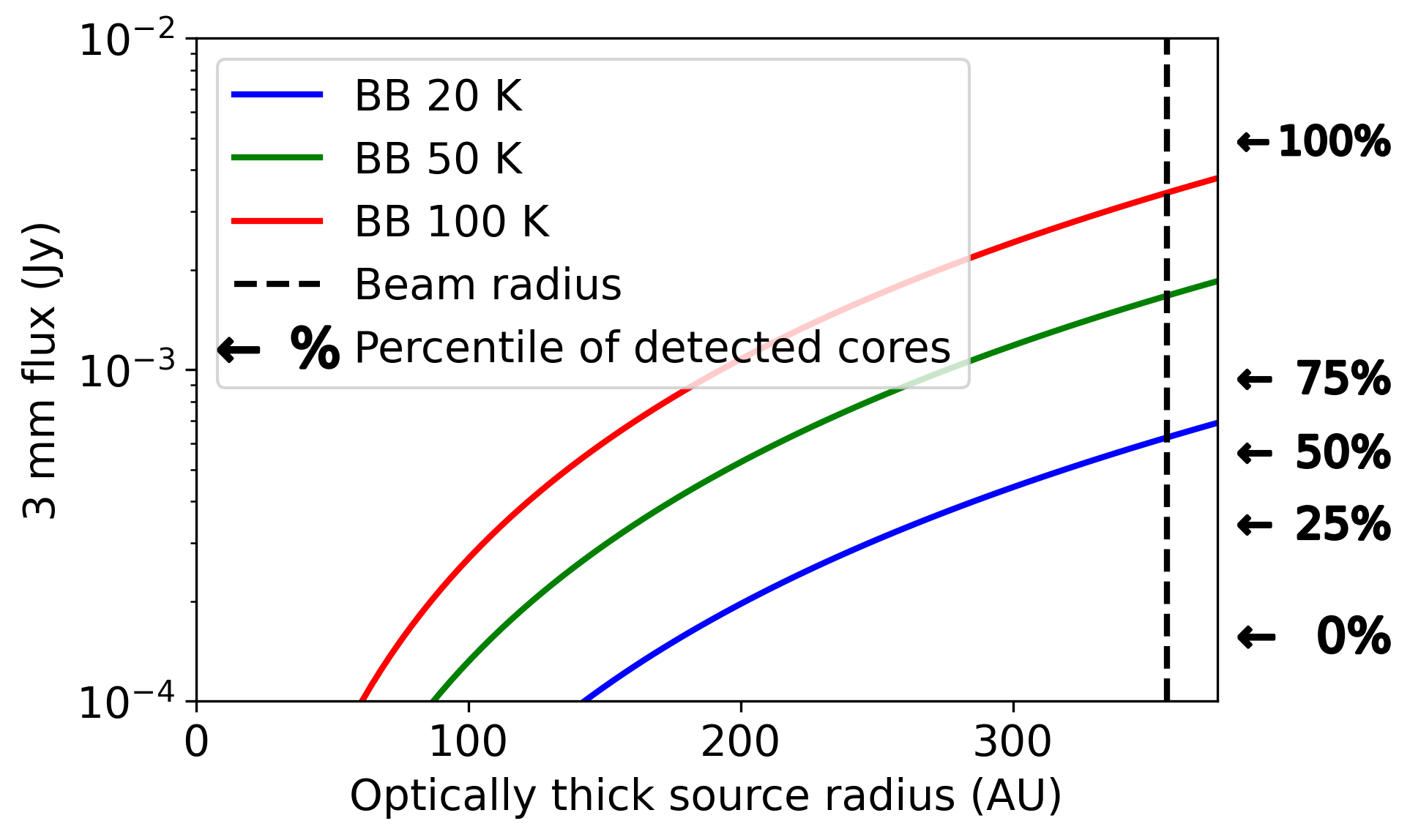}{\columnwidth}{(b)}    
\caption{A model of an optically thick (blackbody) source within our beam at 1 mm (a) and 3 mm (b). By changing the size of the optically thick sphere we can see how many of the sources in our core catalog can be explained using this model. 
Assuming a beam-sized or larger optically thick core, less than 20\% of the 1 mm detections can be explained by core temperatures of 20 K. If some of the sources are smaller and are not completely optically thick, the average temperature would have to be even higher. Thus, we assume a higher average core temperature of 50 K. }
\label{fig:BB}
\end{figure}

If we assume, very simplistically, that all of our observed cores are optically thick beam-sized spheres, about 50\% of the cores at 3 mm and fewer than 20\% of the cores at 1 mm \refreport{can have $T_{KE} < 20$ K based on the expected and observed fluxes}, as shown in Figure \ref{fig:BB}. In reality, it is extremely unlikely that all of the cores are close to beam-sized. Sources smaller than the beam size would need to be warmer to produce the same flux. Furthermore, Figure \ref{fig:SI} shows that most of the cores are not completely optically thick, which requires the assumed temperatures to be even higher. About 75\% of 1 mm detections are consistent with average temperatures of at least 50 K with the remaining 25\% being hotter.
Thus, in this work we take the average core temperature to be 50 K.

To test our assumptions, we create a simple model of our observed sources. It consists of an optically thick sphere of radius $R$ that is surrounded by optically thin dust of a certain surface density. This model mimics a face-on optically thick disk embedded in dust and is intended to confirm that we are able to detect dusty disks. Since many of our sources are calculated to be slightly optically thick, we also use this to test the expected spectral indexes for optically thick sources embedded in optically thin dust. 

\begin{figure}[th]
    \centering
    \includegraphics[width=\columnwidth]{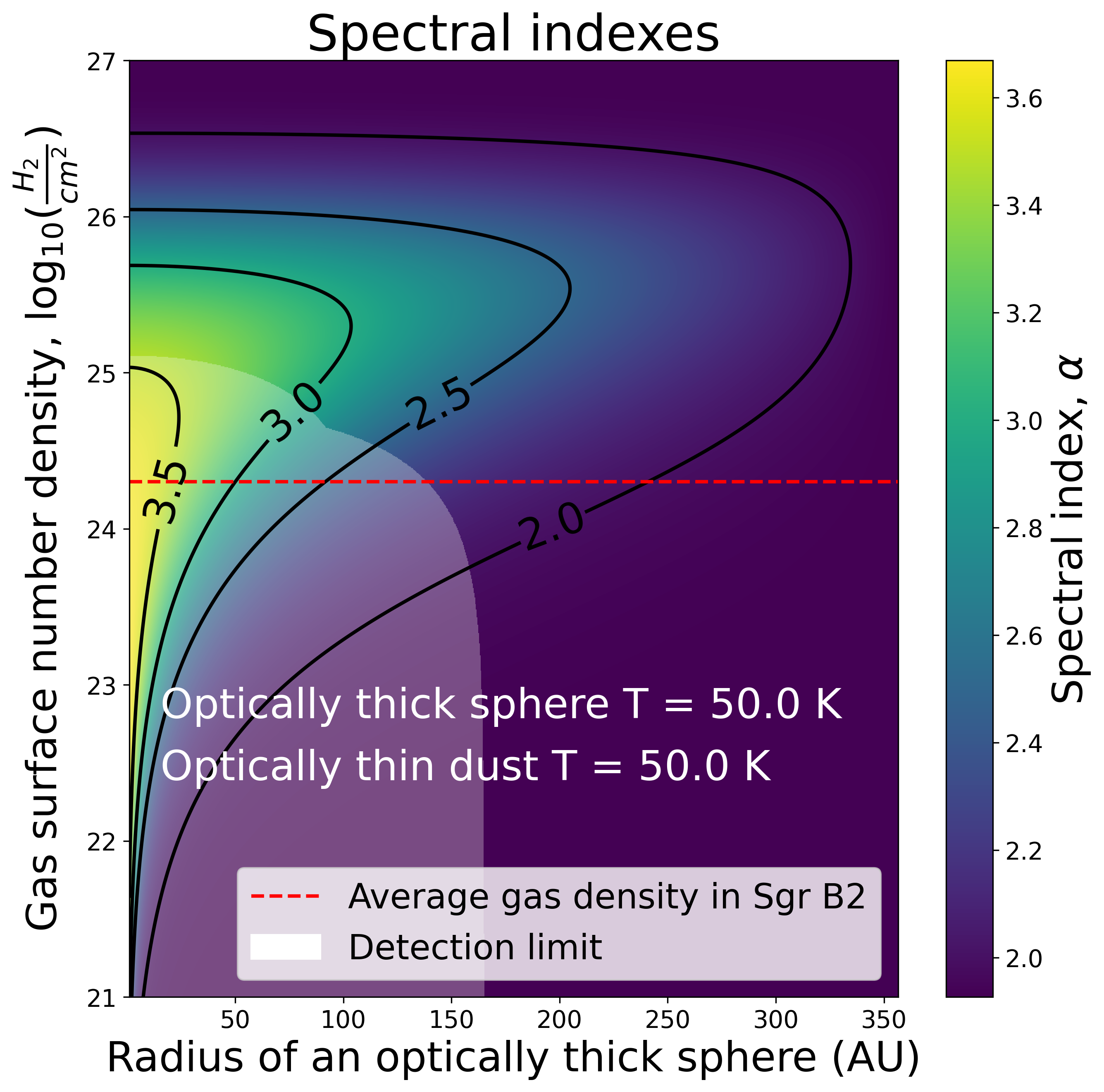}
    \caption{Modeled spectral indexes for an optically thick spherical core of a certain radius surrounded by optically thin isotropic dust of a given gas surface number density. A gas-to-dust ratio of 100 is assumed. The average gas surface density of $3\times10^{24}$ $\mathrm{cm}^{-2}$ in Sgr B2 \citep{Sanchez-Monge2017} is most likely representative of the minimum gas surface density within our beam centered on our detections. We do not expect any sources to be below this line, since all detections represent overdensities. At gas surface densities of $\sim$$4\times10^{26}$ $\mathrm{cm}^{-2}$ the dust surrounding the central sphere becomes optically thick. 
    The spectral index of an optically thick blackbody at 50 K is close to 1.93. The white shaded area indicates a parameter space of sources below our data's sensitivity. }
    \label{alpha_model}
\end{figure}

Assuming a gas-to-dust ratio of 100, we represent the dust surface density using gas column density, with an average over the Sgr B2 N and M cluster of around $3\times10^{24}$ $\mathrm{cm}^{-2}$ \citep{Sanchez-Monge2017}. As discussed above, we assume a dust temperature of 50 K. Figure \ref{alpha_model} shows the expected spectral indexes of such sources as a function of the radius of the central optically thick sphere and the gas surface number density. We further indicate the region on the plot that will not be detectable in either of our bands at our sensitivity as the whited out region on the bottom left. 
The unshaded region of the plot corresponds to the parameters of our simplified model that can explain the observed sources. 

Using this model, for an optically thick face-on disk with an average temperature of 50 K we conclude that:
a) in lower column density regions $(< 10^{24}$ $\mathrm{cm}^{-2})$  we are sensitive to sources that are larger than $\sim$300 AU in diameter and 
b) as the density \refreport{of the surrounding dust} increases past the average density in Sgr B2 (and up to $10^{26}$ $\mathrm{cm}^{-2}$ where the
%removed "surrounding"
dust becomes optically thick) the optically thin dust becomes the dominant source of emission. 

The sources in our catalog are also much brighter than in any of the nearby star-forming regions. In the Orion molecular cloud only four of the brightest disks would be detectable at Sgr B2's distance \citep{Tobin2020}.
Only a few of the disks in Orion are larger than 200 AU with corresponding masses of 0.2\msun. The sources in our catalog are most likely larger, more massive, and warmer than the ones in Orion.

\subsubsection{\hii regions?}
We evaluate whether the sources with $\alpha \approx 2$ could be \hii regions and conclude that they are most likely not.
At brightness temperatures of $10^4$ K, in order to match the range of recorded fluxes in our catalog, an optically thick \hii region would have to be between 30 AU and 80 AU in diameter. 
Using a Str{\"o}mgren sphere as an approximation, we know that:
\begin{equation}
    Q(H^0) = \frac{4}{3}\pi R^3 N (H)^2\alpha_B,
\end{equation}
where $Q(H^0)$ is the number of ionizing photons, $R$ is the radius of the ionized sphere, $N(H)$ is hydrogen number density, and $\alpha_B$ is the recombination rate to all levels excluding the ground level, \refreport{$\approx 3 \times 10^{-13}$ cm$^3$ s$^{-1}$} \citep{Draine2011}. Taking $Q(H^0)$ between $5\times10^{47}$ s$^{-1}$ and $1.5\times10^{49}$ s$^{-1}$, equivalent to B0-O6 stars, we find that 30-80 AU, hyper-compact \hii regions can potentially exist in densities between $10^{7}$ and $10^{9} $ cm$^{-3}$. These densities are reasonable for Sgr B2 \citep{Meng2022}.
However, any optically thick \hii region larger than 80 AU in diameter should be detected with our sensitivity. No such sources are present in our catalog.
Roughly 100 sources in our catalog have $\alpha \approx 2$. 
By making several assumptions, such as higher \hii region metallicity (which will lower its electron temperature) and the right amount of high-mass stars embedded in higher-than-previously-measured gas densities, many of the observed optically thick sources can be attributed to 30-80 AU \hii regions. At the same time, there exists a population of \hii regions larger than 500 AU \citep[e.g. this work,][]{Meng2022}. Assuming that the \hii regions evolve continuously, the absence of 80-500 AU scale \hii regions indicate that there was an abrupt period with no stars being formed. A short, temporary, and complete interruption of star-forming processes would be extremely unusual.
Thus, we claim that there are very few to none \hii regions contaminating our catalog. Additional observations, e.g.\ radio recombination lines, could help determine the nature of these sources definitively.

\subsubsection{Prestellar cores?}
As described in Section \ref{sec:source_masses}, we can place a lower limit on the masses of our sources by assuming each is a beam-sized uniform sphere with a dust optical depth of 1. 
Free-fall time of a spherically-symmetric core is:
\begin{equation}
    t_{ff} = \sqrt{\frac{3\pi}{32G\rho}} = \sqrt{\frac{3\pi V}{32Gm}} = \sqrt{\frac{\pi^2 R^3}{8Gm}},
\end{equation}
where $G$ is the gravitational constant and $V$, $m$, and $R$ are the volume, mass, and radius of a core. 
Taking the least massive source in our catalog ($\sim$1\msun) to be beam-sized ($\sim$700 AU) would result in free-fall times of 1200 years, while the most massive core of the same size would have a free-fall time of 40 years -- extremely short on star formation scales.
Thus, it is unlikely that any of our sources (at least on $\sim$700 AU scales) are still undergoing free-fall collapse, indicating that there are no prestellar cores in our sample. 
While this is an extreme approximation, the higher mass, heterogeneous nature of the source, and the influence of the surrounding environment would speed up the infall of the material \refreport{and result in even shorter collapse timescales}. 

We can further estimate the effects of the high-pressure environment using the Bonnor-Ebert mass:
\begin{equation}
    M_{BE} = 1.18 \frac{c_s^4}{G^{3/2}p_{BE}^{1/2}},
\end{equation}
where $c_s$ is the speed of sound, 0.4 km/s (T = 50 K), $G$ is the gravitational constant, and $p_{BE}$ is the pressure ($P/k_B = 10^{8} - 10^{13} \mathrm{\ K\ cm}^{-3}$ \citep{Myers2022}).
The maximum Bonnor-Ebert mass for the sources in Sgr B2 is then \refreport{$\sim$$10^{-3}$$-10^0$\msun}. Only two cores in our catalog have mass lower limits below 1\msun. Thus, the majority of the cores are not pressure supported.

\refreport{Presently, there is not enough observational evidence to determine the impact of magnetic fields in the collapse of massive cores \citep{Pattle2023}. The observed randomness of the core's magnetic field orientation relative to the large-scale magnetic fields in Cygnus X suggests that on small scales, gravity plays a more important role compared to magnetic field \citep{Ching2017}. Thus, in this work we assume that the magnetic fields do not contribute to the core collapse support.}
Having ruled out the free-fall collapse and pressure support, and the improbable magnetic field support, the remaining option is the rotational support of disks.

\subsubsection{Stage II and III YSOs?}
\refreport{To test whether the stars are Stage II YSOs, we check what stellar mass would be predicted by extrapolating local relations.}
%With a dust mass lower limit of the cataloged sources, we can approximate the masses of the central stars for Stage II YSOs. 
We extrapolate the ratio of the dust mass and the central star of $\sim$1\% \citep{Williams2011} one order of magnitude for more massive objects to calculate the possible mass of the central star for the least massive sources in our catalog. 
The faintest source in our data is 1.2 mJy at 1.3 mm, which would correspond to a central star with a mass of at least 30\msun. 
If all of our observed sources are Stage II YSOs, the median star mass would be around 300\msun, which is on the order of the largest stars ever detected.

Current disk demographic samples \citep[e.g.,][]{Andrews2018, Williams2011} are limited to M $\lesssim2$\msun. 
% removed "Finally,"
If we assume an optically thick, face-on dusty disk with an average temperature of 50 K, the disk would have to be at least 200 AU in size to be detectable with our sensitivity\refreport{, which is similar in size to the largest disks in the Solar neighborhood.}
All of our targets, if they are disks, must be larger and more massive than any disks observed in the local neighborhood \citep[e.g.,][]{Andrews2018}.
Stage III YSOs do not have infrared excess and thus could not possibly be detected in our data.

\subsection{Inferred stellar mass} \label{sec:inferred_mass}
\refreport{
We estimate the total stellar mass of Sgr B2 N and M using direct source counting down to 1\msun. We take the calculated masses from Section \ref{sec:source_masses} and extrapolate the low-mass end of the Kroupa initial mass function \citep{Kroupa2001} to obtain the core-based inferred stellar mass. We count the known \hii regions to add the high-mass sources.  We report the total observed ($M_{\Sigma obs}$) and inferred ($M_{\Sigma inf}$) stellar masses for Sgr B2 N and M in Table \ref{tab:inferred_mass}. 
}

\refreport{
Using individual core masses provides more accurate mass estimation, compared to lower-resolution studies, where an average core mass is assigned.
To allow for direct comparison with previous works, we take the radii of Sgr B2 N and M clusters to be 0.4 pc and 0.5 pc, respectively. We further assume that roughly one third of a core mass will end up in a star \citep{Lada2003}. We take 6$\sigma$ as our completeness limit within the cluster boundaries -- 0.3 mJy for Sgr B2 M and 0.2 mJy for Sgr B2 N, which corresponds to 1.5\msun and 1\msun stellar mass respectively (equivalent to 4.5\msun and 3\msun core mass). We exclude all the sources below this completeness limit.
Since the massive cores rapidly evolve into \hii regions, we also set an upper limit of 20\msun. Thus:
}
\begin{equation}
    M_{core,inf} = M_{core,obs} \frac{\int_{0.3\msun}^{120\msun}f \mathrm{d}m}{\int_{6\sigma}^{20\msun}f \mathrm{d}m},
\end{equation}
\refreport{where $M_{core,inf}$ is the inferred core-based stellar mass, $M_{core,obs}$ is the sum of the core masses above our completeness limit divided by 3, and $f$ is the Kroupa IMF.
Compared to G18 we detect $\sim$10 times more cores in Sgr B2 N and $\sim$4 times more cores in Sgr B2 M. With the larger number of sources and augmented stellar mass calculation, we observe a 60\% higher core-based stellar mass in Sgr B2 N, but 30\% lower in Sgr B2 M.}

\refreport{
To account for the high-mass sources, we count the number of \hii regions from \cite{DePree1998} and \cite{Meng2022} and add the newly discovered \hii regions in this work. Assuming that each \hii region is powered by a $>$ 20\msun star, the IMF-weighted average mass of such star is 45\msun. Combining the core-based stellar mass with the \hii region-based stellar mass:}
\begin{equation}
    M_{\Sigma inf} = \left( M_{core,obs} + M_{\hii,obs} \right) \frac{\int_{0.3\msun}^{120\msun}f \mathrm{d}m}{\int_{6\sigma}^{120\msun}f \mathrm{d}m},
\end{equation}
\refreport{where $M_{\Sigma inf}$ is the total inferred stellar mass and $M_{\hii,obs} = N(\hii) \times 45.5 \msun$.
Taking the star-forming age of the cloud as 0.74 Myr \citep{Kruijssen2015}, we estimate the star formation rate (SFR) for each cluster. Compared to G18 we find that the total inferred stellar mass in Sgr B2 N increased by more than a factor of two, but is slightly lower for Sgr B2 M.}

\begin{table*}[th]
\centering
\caption{\refreport{Inferred stellar mass of Sgr B2 N and M using \refreport{direct} source counting. \refreport{The clusters' boundaries are chosen to be consistent with previous works: R = 0.4 pc and R = 0.5 pc for Sgr B2 N and M respectively.} 
Using the calculated core mass \refreport{from Section \ref{sec:source_masses} and assuming a 30\% star formation efficiency}, we obtain the \refreport{observed} core-based stellar mass (\refreport{$M_{core,obs}$}). 
We assume a lower completeness limit to be $6\sigma$ ($\sim$1.5\msun) and the upper lifetime-based limit of 20\msun in stellar mass and use Kroupa IMF \citep{Kroupa2001} to infer the core-based stellar mass ($M_{core,inf}$). 
We count the number of \hii regions, $N(\hii$), from \cite{DePree1998}, \cite{Meng2022}, and this work and assume that each \hii region contains a 45.5\msun star to populate the M $>$ 20\msun tail of the IMF. Combining the direct counting of cores and \hii regions, we present the total observed stellar mass ($M_{\Sigma obs}$) and the total inferred stellar mass ($M_{\Sigma inf}$) in each cluster. We include the results from the low-resolution observations (G18) and \hii region modeling work \citep{Schmiedeke2016}. The star formation rate assumes $t_{\mathrm{SF}} = 0.74$ Myr.}}

\begin{tabular}{ccccccccccc}
\hline
\hline
Name     & $N(core)$ & $M_{core,obs}$ & $M_{core,inf}$ & $M_{core,inf}^{G18}$ & $N(\hii)$ & $M_{\Sigma obs}$ & $M_{\Sigma inf}$ & $M_{\Sigma inf}^{G18}$ & $M_{\Sigma inf}^{S16}$ & SFR \\
         &  & \msun & \msun & \msun &  & \msun & \msun & \msun & \msun & \msun yr$^{-1}$ \\

\hline
N & 105      & 1010          & 2400               & 1500         & 11     & 1500        & 2800         & 1200   & 2400   &   0.0038   \\
M & 68       & 550           & 1600               & 2300         & 57     & 3100        & 6900         & 8800   & 20700  &  0.0093    \\ \hline

\end{tabular}%

\label{tab:inferred_mass}
\end{table*}

With the high-resolution data, it is evident that the historically defined sizes of Sgr B2 \refreport{North and Main} in \cite{Goldsmith1990} do not necessarily represent which sources are part of the clusters. Both N and M have a continuous ``chain" of sources in projection in the eastward direction, as visible in Figures \ref{fig:B3} and \ref{fig:B6}. The defined sizes cut through these ``chains" of sources.
Further kinematics studies of the region that extended beyond the centers of the clusters \citep[e.g.][]{Schworer2019} are needed to determine which sources belong to which protocluster and would allow for more accurate count-based cluster mass estimation.

% removed a bunch about IMF slope

\subsection{Core fragmentation} \label{sec:core_fragmentation}
We compare our core catalog with G18, which observed Sgr B2 N and M at 10 times coarser resolution. For each low-resolution detection, we find the number of high-resolution detections within 1.5 $\times$ beam FWHM of the source center. 
We exclude sources from the G18 catalog that would not be detectable in our data, specifically at the pointing edges where the sensitivity is lower. \refreport{Of the 94 potentially detectable G18 sources, only 77 are detected at high-resolution.} 
Out of 371 detections at 3 mm in our catalog, 109 cores are contained within 
% removed "the fragmented"
low-resolution sources, 42 are one-to-one match, and the remaining 220 are new detections with no corresponding detections in G18.

\refreport{Of the G18 sources that are not detected in the high-resolution data, some are most likely false detections, but some may be resolved-out extended structures.}
Our shortest baseline length of 41 m is longer than the 15 m in G18 data. On average, our images are made weighting longer baselines. As \refreport{a}
% the -> a
result, some of the previous detections \refreport{may}
% can -> may
be resolved out. We find that over a dozen sources from the low-resolution catalog do not have a counterpart in the high-resolution catalog.
The high-resolution data's largest recoverable scale is 0.9", or $\sim$7500 AU. 
% removed: While some faint and extended sources might indeed have been resolved out, at least some of the low-resolution detections are likely false positives due to imaging artifacts. 

\begin{figure}[th]
    \centering
    \includegraphics[width=\columnwidth]{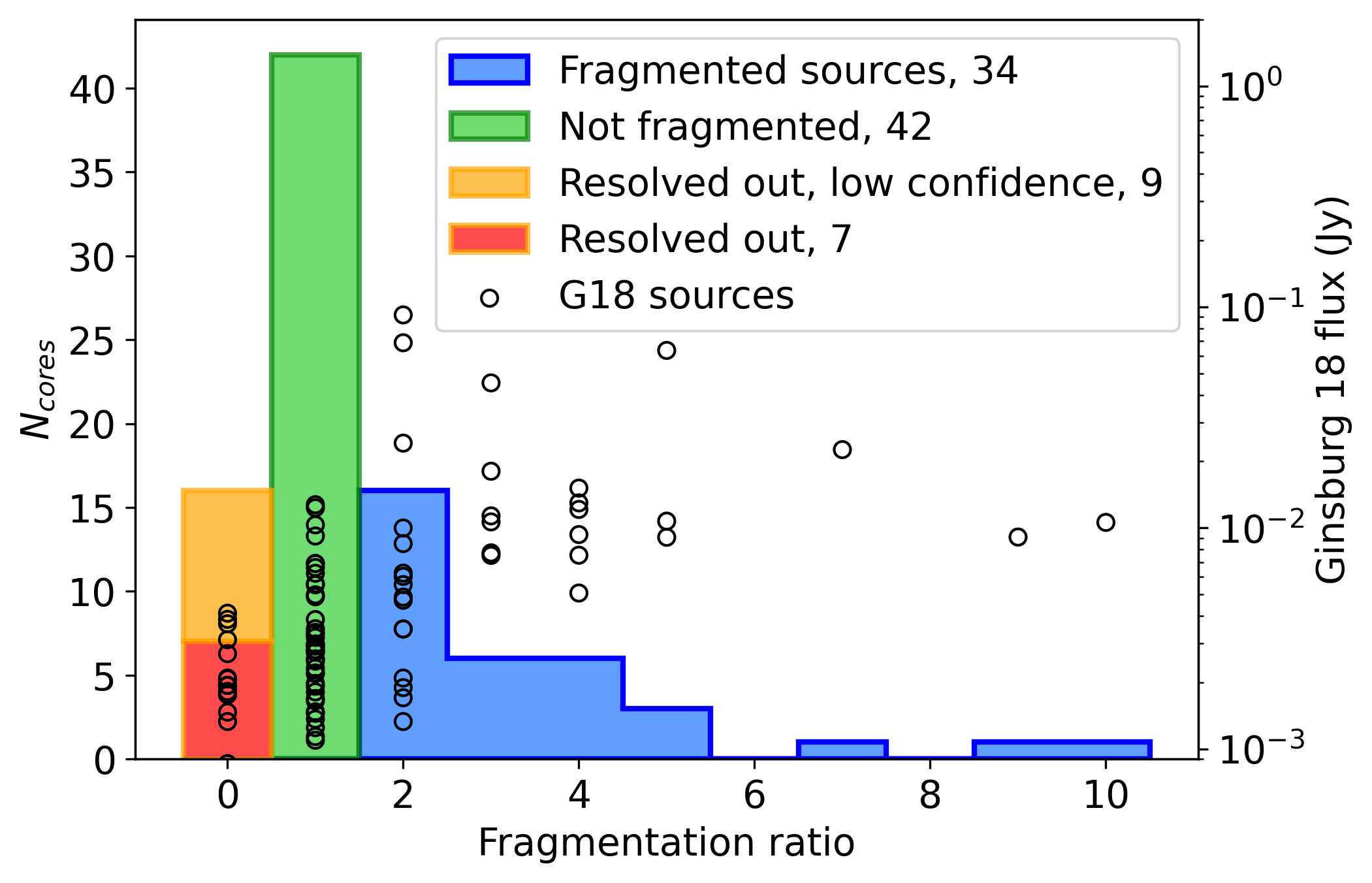}
    \caption{Number of sources from our data that fall within 1.5 $\times$ FWHM of G18 sources. Over 50\% of the G18 sources fragment into two or more sources. The black circles show the relationship between the flux and the fragmentation ratio of G18 sources.
    The left-most bin shows sources from G18 that do not have a counterpart in our data. On average, each G18 source fragments into two.} 
    \label{fig:highperlow}
\end{figure}

\begin{figure*}[ht]
    \centering
    \includegraphics[width=0.95\textwidth]{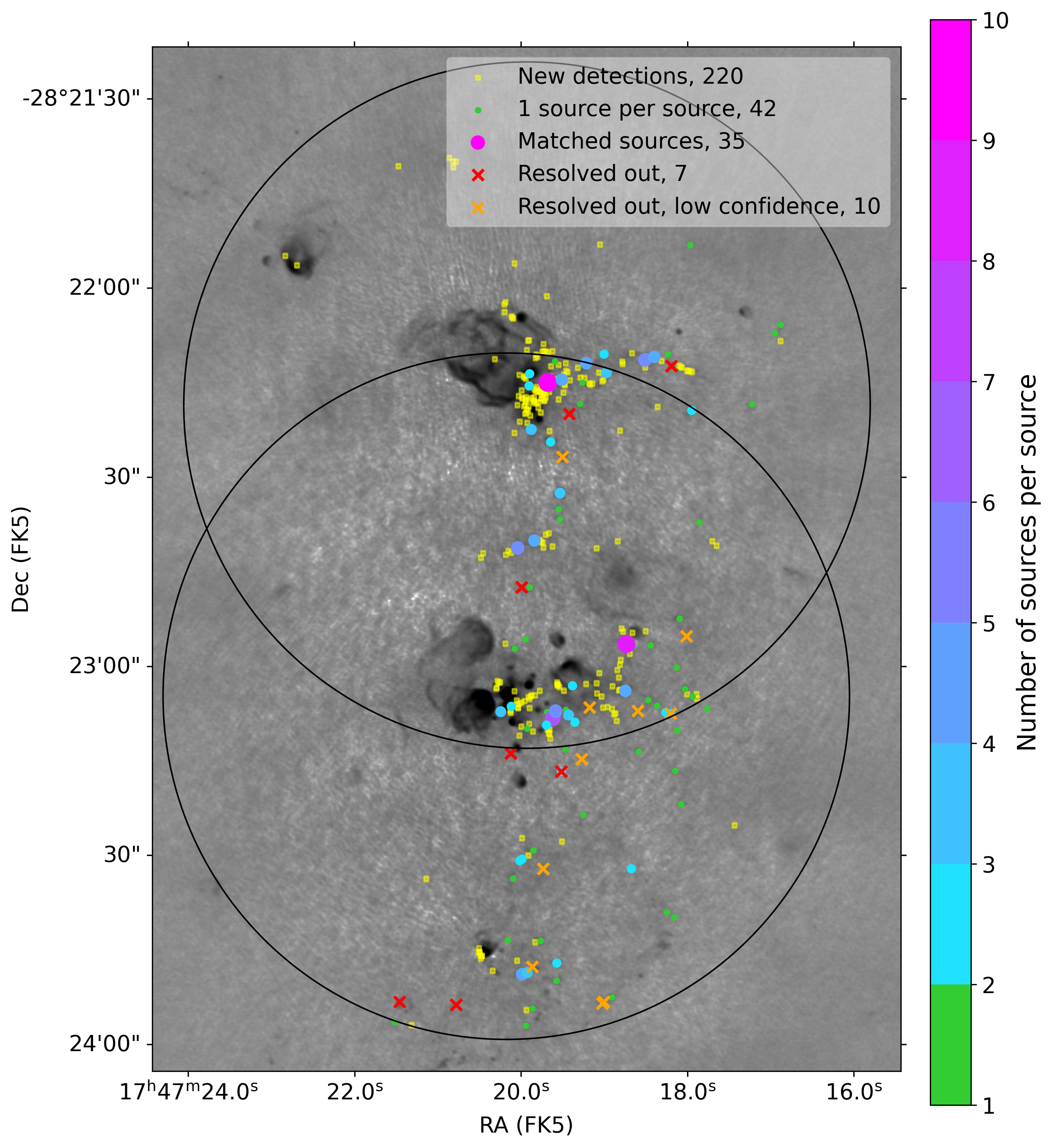}%{Overview_sps.png}
    \caption{3 mm ALMA mosaic of Sgr B2 adapted from \refreport{G18}. The black circles show the field of view of the two 3 mm pointings in this work. The filled-in circles show the sources that have been matched by-hand between the two data sets, with \hii regions disregarded. The color and the size of the circles show the number of higher-resolution sources within 1.5 $\times$ FWHM of the lower resolution beam, which we interpret as fragmentation. The x's indicate sources resolved out with the increase in resolution and increase in the shortest baseline length. ``Low confidence" detections from \refreport{G18} are shown in orange. The detections visible in the figure but not marked are either \hii regions or are located at the edge of the high-resolution pointings where the sensitivity was below the detection threshold for such sources. Some of the ``resolved out" sources (including the red high-confidence detections) are likely to be false detections caused by imaging artifacts.} 
    \label{fig:comparison}
\end{figure*}

We find that over half of the low resolution detections ``fragment" into two or more sources, with one of the sources fragmenting into ten, as shown in Figure \ref{fig:zoomin}. 
The distribution is shown in Figure \ref{fig:highperlow}. We do not find any pattern in the locations of the cores that fragment vs those that do not. Figure \ref{fig:comparison} shows the spatial distribution of the cross-matched and resolved-out sources as well as the new detections.

On average, each detected G18 source fragments into 2 cores. If we take the one third of the sources that fragment, then there is an average of 3.4 cores per G18 fragmenting source. In general, source fragmentation changes the shape of the core mass function (and if these fragments represent individual systems -- the IMF), making it \refreport{steeper}.
%removed "less steep"
Because cores previously inferred to contain single, very massive star are broken into many lower-mass objects, increasing the total number of sources, the total expected luminosity \refreport{of these sources} reduces significantly.

We find that any source from G18's catalog, at 5000 AU resolution, with flux above 20 mJy ended up being fragmented at 700 AU resolution. It is not possible to place meaningful lower limits on the fragmentation function at low flux: the sources are more likely to be close to the sensitivity limit of the higher-resolution data.

\section{Summary} \label{sec:sum}
We presented $\sim$700 AU follow-up continuum observations of a massive molecular cloud in the CMZ -- Sagittarius B2. We detected \refreport{410} unique cores: 371 at 3 mm, a twofold increase compared to \cite{Ginsburg2018}, and \refreport{218} at 1 mm. Most of the newly detected sources are due to the increased sensitivity, while about a fifth come from resolving fragmented sources. We find most sources to be marginally resolved and have a spectral index consistent with ${2<\alpha\lesssim3}$. 
We infer that the dusty objects we observe must be 
\refreport{disks that are warmer than those in the rest of the Galaxy}. 
% old: "warmer than disks observed in the Galaxy"
Because of their high temperatures, these sources are unlikely to be prestellar cores. Because of their high column densities, they are unlikely to be Stage II or later YSOs. 
We therefore conclude that the sample is comprised primarily of Stage 0/I YSOs. \refreport{We find that the core mass function of our sample is consistent with the Kroupa IMF.} Using direct source counting, we infer the total stellar mass of \refreport{2800\msun} for Sgr B2 N and \refreport{6900\msun} for Sgr B2 M. The resulting star formation rate is \refreport{0.0038\msun yr$^{-1}$ for Sgr B2 N and 0.0093\msun yr$^{-1}$ for Sgr B2 M totalling 0.013\msun yr$^{-1}$, which is similar to previous estimates}.

\section*{Acknowledgments}
This paper makes use of the following ALMA data: ADS/JAO.ALMA\#2016.1.00550.S, ADS/JAO.ALMA\#2013.1.00269.S. ALMA is a partnership of ESO (representing its member states), NSF (USA) and NINS (Japan), together with NRC (Canada), MOST and ASIAA (Taiwan), and KASI (Republic of Korea), in cooperation with the Republic of Chile. The Joint ALMA Observatory is operated by ESO, AUI/NRAO and NAOJ. The National Radio Astronomy Observatory is a facility of the National Science Foundation operated under cooperative agreement by Associated Universities, Inc.
AG acknowledges support from the NSF under CAREER 2142300 and AST 2008101 and 2206511.
NB and DJ were supported by NRAO Student Observing Support grants SOSPA8-010 and SOSPA6-026.

\vspace{5mm}
\facilities{ALMA}

\software{Image reduction and final catalogs are available on GitHub: \url{https://github.com/nbudaiev/SgrB2_ALMA_continuum/releases/tag/Accepted-15-Sep-2023}. This work utilized the following tools: \texttt{astropy} \citep{astropy:2013, astropy:2018}, CASA version 5.7.0-134.el7 \citep{McMullin2007}, \texttt{astrodendro} \citep{Rosolowsky2008}, \texttt{plfit} (\url{https://github.com/keflavich/plfit})}, \texttt{dendrocat}, \texttt{radiobeam}.

\setlength{\bibsep}{0.0pt}
\bibliographystyle{aasjournal}
\bibliography{references}

\appendix
\refreport{
\section{Line contribution to continuum}\label{sec:line_sub}
It has been shown that sources towards the central subclusters Sgr B2 N and M are line-rich \citep{Sanchez-Monge2017}. We inspected the full cubes of Sgr B2 N Band 3 and Band 6 pointings to evaluate the relative line contribution to the sources in our catalog.
We extracted spectra at the location of the brightest pixels of the cores cataloged in this work.
We removed line emission and absorption from the spectra using the \texttt{astropy's} implementation of sigma clipping with \texttt{sigma}=2 and \texttt{maxiters}=5. First, we calculated the mean of spectra before and after line removal for each core. Then, to estimate the proportional flux change, we calculate the difference between the two spectra and divide by the brightest pixel's flux in the non-line-subtracted continuum image used to generate the catalogs. We find that the mean relative line contribution for all cores is $-4.4\%$ for Band 3 and $-1.2\%$ for Band 6. The scaled mean average deviation is 5.2\% for Band 3 and 11.2\% for Band 6. All sources with significant flux change due to line subtraction had SNR $<$ 10.
Thus, we conclude that on average emission and absorption lines do not contribute to the continuum level significantly for the cataloged objects. 
}

\section{Cleaning and self-calibration testing}\label{sec:testing}
Sgr B2 region is one of the most complex regions that ALMA can observe. Pointing Band 3 beam at one of the two largest cores N or M places the other of the two just on the edge of the beam, significantly contributing to the artifacts in the image. We tested different cleaning approaches with special focus on masking and choosing the right cleaning threshold.

\begin{figure}[th]
    \centering
    \fig{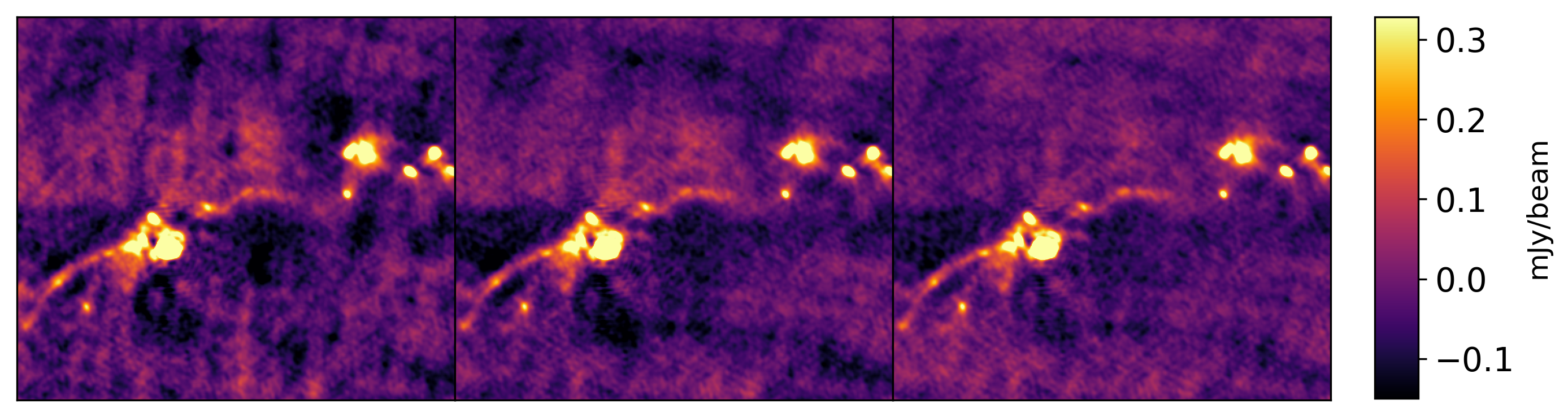}{0.95\columnwidth}{(a) Sgr B2 Z10.24 \hii region and the surrounding cores from the Sgr B2 N pointing at 3 mm. The bright Sgr B2 N at the center of the pointing is causing vertical streaking in this region. The bright \hii region shown in the cutout is causing concentric ring-like artifacts. \refreport{The pointing is centered on ICRS 17:47:19.931; -28:22:40.737.}}
    \fig{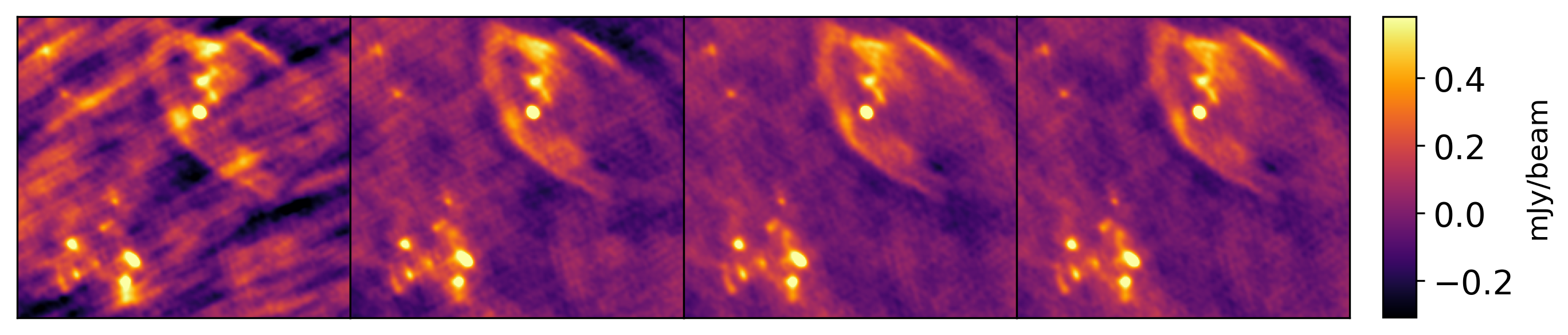}{0.95\columnwidth}{(b) Sgr B2 Y \hii region and the surrounding cores from the Sgr B2 M pointing at 3 mm. The radial streaks are caused by the extremely bright source at the center of the pointing -- Sgr B2 M. \refreport{The pointing is centered on ICRS 17:47:18.683; -28:22:55.726.}}
    \fig{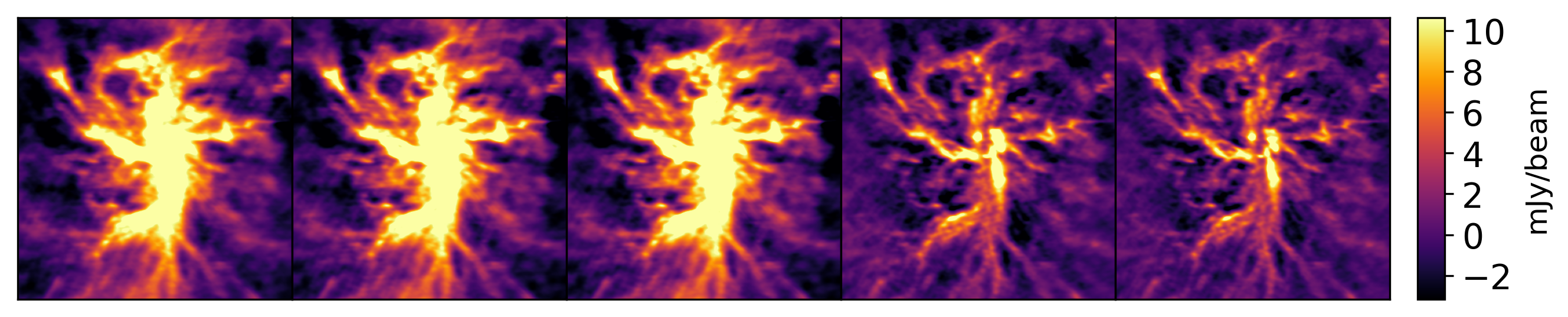}{0.95\columnwidth}{(c) Sgr B2 N at 1 mm. Diagonal artifacts are mitigated throughout the self-calibration process. The first iteration of the amplitude self-calibration (4th panel) removed large-scale ripples centered on the cutout causing a perceived reduction in flux. \refreport{The pointing is centered on ICRS 17:47:19.874; -28:22:18.604.}}
    \fig{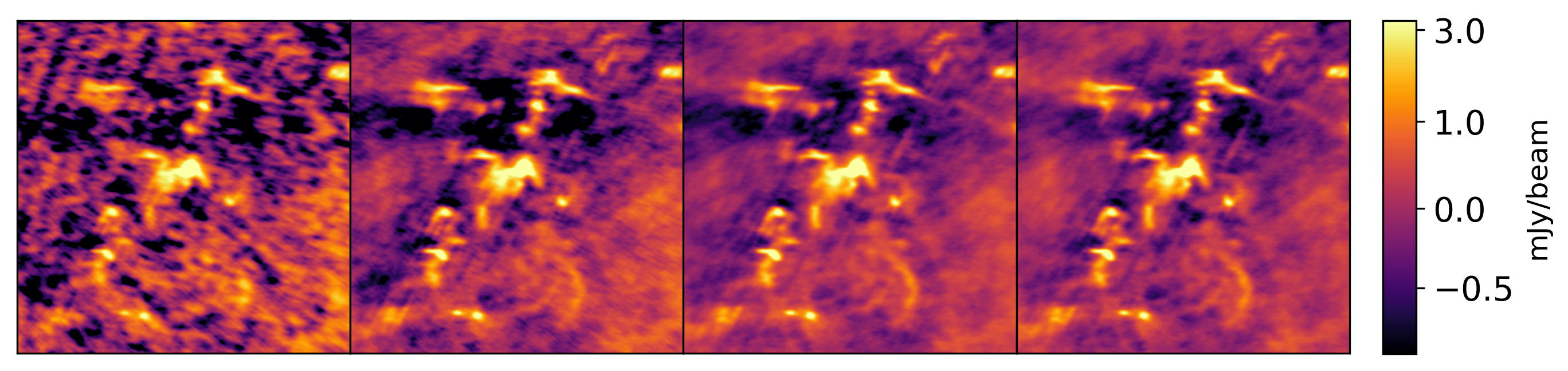}{0.95\columnwidth}{(d) A clump of sources South-East of Sgr B2 M at 1 mm. \refreport{The pointing is centered on ICRS 17:47:19.631; -28:23:08.542.}}

\caption{A sample of \refreport{$\sim$$0.4 \times 0.4 $ pc} zoom-ins for each pointing at different calibration stages highlighting different types of artifacts. The most prominent artifacts, concentric rings and radial streaks, are caused by the extremely bright sources in the field-of-view. 
The iterations from left to right, if present, are: uncalibrated, phase, \texttt{solint} = \texttt{inf}; phase, \texttt{solint} = \texttt{int}; phase and amplitude, \texttt{solint} = \texttt{inf}; phase and amplitude, \texttt{solint} = \texttt{15s}.}
\label{fig:calibration}
\end{figure}

We found that the choice of mask significantly affects the quality of the final image. 
Generally, it is recommended to mask the images conscientiously, only including the real emission \citep{Richards2022}. At first, we tried to use RMS-based masks. We went through the calibration process using $5\sigma$, $3\sigma$, and $2\sigma$ masks based on the noise in the inner, signal-free region of the image. The resulting residuals contained very sharp edges around the mask contours. We then switched to a less restrictive approach: we drew polygons around visually identifie\refreport{d} emission regions. The resulting mask covered more emission, but also included some of the artifact-rich regions. After cleaning, the images ended up with several factors lower noise than with RMS-based masks and the residual did not have any abrupt features. To minimize the effect of including the artifacts in the cleaning mask, especially in the first few iterations where they are the most prominent, we decided to use the ``strict" mask for the first two iterations and then use the more broad mask for the remainder of the cleaning and calibration process. This time, we drew two sets of masks for each pointing: one containing just the brightest emission similar to the 5$\sigma$ mask, but also including significant sources of emission further away from the center of the image, and a more broad mask where polygons surround any regions with visible emission. The use of a combination of a strict and broad masks resulted in a few percent lower RMS than using only the broad mask.

Generally, we used $6\sigma$ cleaning threshold for the first two cleaning iterations and $3\sigma$ for the remainder of the cleaning and calibration. However, the Band 6 pointing of Sgr B2 N suffered from divergence at these thresholds. Using higher-threshold cleaning, we determined that the image significantly improves after amplitude and phase calibration. After testing different cleaning thresholds, we found that the deepest we can clean the image without the clean diverging is $\sim$$10\sigma$ until the amplitude and phase calibration and then the 3$\sigma$ cleans will not diverge. The lowest measured RMS for each pointing is presented in Table \ref{tab:RMS}. The self-calibration progress images highlighting different regions are shown in Figure \ref{fig:calibration}.

\refreport{
\subsection{Intrinsic uncertainty of interferometric imaging}
The choice of parameters during imaging and self-calibration can have a significant influence on the final result, especially for bright, extended sources. 
We explored possible approaches to create a more robust version of the final image and evaluated the possible systematic and random uncertainties introduced in the process. We used the Band 6 Sgr B2 N pointing for the testing and compared the brightest pixel for each of our cataloged source between different versions of the image. The observations for each pointing were split into two measurement sets that were observed on different days; we were modifying each measurement set separately and combining them for the final comparison.
}

\refreport{
\subsubsection{Problematic baselines}
As visible on Figure \ref{fig:B3} and especially on Figure \ref{fig:B6}, there are large-scale parallel ripples. The impact of these artifacts is especially visible in Figure \ref{fig:calibration}(c) between panels 3 and 4, where the artifacts become less prominent during amplitude self-calibration. A possible approach to remove these artifacts is to identify and flag the problematic baselines that are causing the ripples. Using the separation between the wave peaks and the relative orientation of the ripples, we iteratively removed problematic baselines until further removal resulted in diminishing returns. We removed ten most severe baselines.
Since the removed baselines were all very short baselines, we also lost the large-scale structures. The removal of individual baselines had very little effect on self-calibration solutions. 
Removing short baselines introduced individual source flux scatter of up to 50\%, primarily for dim sources, but no systematic flux changes were present.
}

\refreport{
\subsubsection{QA-flagged data}
We discovered that one of the measurement sets for Sgr B2 N Band 6 pointing had 16 ID-sequential antennas flagged during the very first stages of ALMA pipeline run prior to the delivery to the user. The flagging starts about two-thirds into the observing run, right after the water vapor calibration scan. Based on the limited information, we attribute this flagging to a fault in the correlator. Suddenly dropping almost a third of antennas results in a discontinuity in amplitude self-calibration solutions for some antennas. We chose not to flag the rest of the data past the time when the partial flagging begins, as it would result in losing over ten percent of observations. Instead, we attempt to improve the amplitude self-calibration solutions as described in the next section.
}

\refreport{
\subsubsection{Poor self-calibration solutions}
We found that excluding solutions from short baselines (that usually have the largest unweighted amplitudes) would result in smaller amount of discontinuities in amplitude self-calibration solutions. We used \texttt{uvrange} parameter in 
\texttt{gaincal} task to limit the baseline lengths used to calculate the calibration solutions. After testing different values, we found that setting \texttt{uvrange} at $> 240$ m results in the least amount of discontinuities in the solutions. 
To further mitigate the possible erroneous flux changes from the amplitude self-calibration, we chose not to apply any antenna solutions that have discontinuities or sharp changes.
}

\refreport{
\subsubsection{Final result}
We combined the approaches described in the subsections above to create as conservative image as possible. We started with the same phase self-calibration process as the original reduction with the exception of excluding the problematic baselines. During the first round of amplitude self-calibration we used \texttt{uvrange > 240m} in \texttt{gaincal} task for the measurement set with flagged data. In addition, we normalized the solutions and increased the solution SNR requirement from 3 to 5 compared to the original imaging parameters. We then inspected the amplitude self-calibration solutions, and very conservatively excluded any solution that might lead to erroneous corrections. Specifically, if there was a discontinuity of more than 0.15 in amplitude or rapid changes of more than 0.5 in amplitude, the corresponding antenna solutions were flagged. We then repeat the same process, now with \texttt{solint} set to 15 s. We then compared the brightest pixel in each cataloged source between the originally reduced data and using the conservative approach. The flux ratios are shown in Figure \ref{fig:reduction_tests}. We find that there is a scatter in flux ratios of up to 40\%, especially for dim sources and the average ratio is very close to 1. This indicates that we can expect a large flux uncertainty for individual sources due to radiointerferometric imaging, but there is no systematic error in the approach we used to create the final catalogs, at least when comparing to a ``more robust" approach. With the self-calibration approach used in this paper, we are able to achieve 20\% lower noise compared to the more conservative self-calibration.
}

\refreport{
\begin{figure}[ht]
    \centering
    \includegraphics[width=0.5\columnwidth]{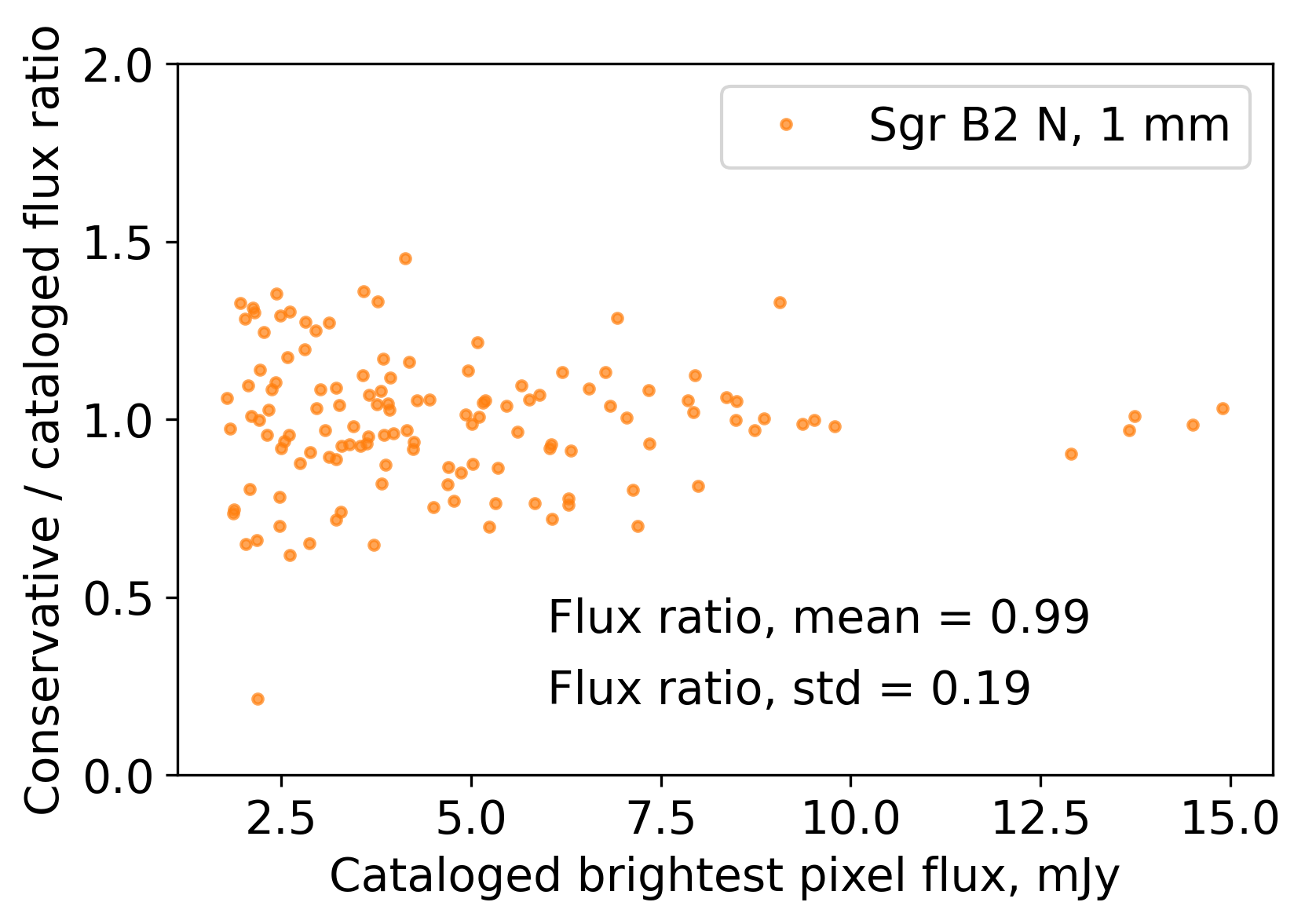}
    \caption{The source peak flux comparison between the images generated with the standard reduction procedure used in this paper and a more conservative approach for Sgr B2 N Band 6. We find that there is no systematic error in the imaging and self-calibration approach we used. However, we see individual flux differences of up to 40\% due to intrinsic uncertainty of radio imaging.}
    \label{fig:reduction_tests}
\end{figure}
}
\begin{deluxetable}{cccccccc}
    \tablecaption{Lowest RMS within the inner, signal-free region of each pointing during different calibration stages. The cleaning thresholds for the first two cleaning iterations were $\sim$$6\sigma$ and $\sim$$3\sigma$ for the remainder of the self-calibration process. The last two columns show the source extraction threshold values within the inner 40\% and 90\% of the imaged area. The noise was calculated within the concentric annuli described in Section \ref{sec:source_extraction}.  \label{tab:RMS}}
    \tablehead{
    Pointing & Uncalibrated     & Phase, inf     & Phase, int & Phase+amp, inf & Phase+amp, 15s    & $4\sigma$-$40\%_{image}$ & $4\sigma$-$90\%_{image}$ \\
     & mJy & mJy & mJy & mJy & mJy & mJy & mJy \\
    }
    \startdata
    N, 3 mm & 0.047 & 0.044 & 0.032 & --    & --   & 0.13        & 0.43        \\
    M, 3 mm & 0.105 & 0.068 & 0.054 & 0.051 & --   & 0.20        & 0.61        \\
    N, 1 mm & 1.37  & 1.35  & 1.35  & 0.47  & 0.35 & 1.41        & 4.62        \\
    M, 1 mm & 0.55  & 0.40  & 0.30  & 0.25  & --   & 1.00        & 3.27       \\
    \enddata
\end{deluxetable}

\refreport{
\subsection{Imaging overlapping pointings} \label{sec:imaging}
At 3 mm (Band 3), pointings centered on Sgr B2 N and M have an overlap in the imaged area, such that the center of one pointing appears at the edge of the other pointing when imaged beyond primary-beam-response of $\sim$0.2. We tested three possible ways to image and reduce these pointings: imaging both pointings together using mosaic gridder, imaging each pointing separately and taking a weighted average of the overlapping region, and imaging and reducing each pointing separately. In this work, we use the latter method where we crop the two imaged pointings based on the achieved noise.
}

\refreport{
We find no significant benefit of imaging the two pointings together with mosaic gridder when compared to using standard gridder on each pointing individually. For these data, the main benefit of mosaic gridder is to make a better model of the image that accounts for more of the bright emission that otherwise might have been outside of the imaged area of a single pointing. In a single pointing, the flux from outside of the imaged area ends up spread over the whole image resulting in higher observed flux. However, we choose the primary-beam-response level of 0.1 such that most of the bright flux, in this case the central clusters N and M, are included in both pointings. When comparing the jointly- and separately-imaged pointings, the reduction in the intensity of artifacts is minimal. Furthermore, the two pointings have a different final cleaning threshold, as shown in Table \ref{tab:RMS}. Mosaicing the two pointings would result in either undercleaning or overcleaning of one of the pointings. Finally, computationally the cleaning of Band 3 mosaic requires close to maximum computing resources available; this would not allow for testing of enough of the imaging parameter space to achieve the desired result.
}

\refreport{
An alternative approach that benefits from the overlapping pointings is to take a weighted average of the separately reduced images. This method works best when the noise is isotropic and we would expect a factor of $\sqrt{2}$ reduction in noise. However, as seen in Figure \ref{fig:B3} interferometric artifacts are highly structured, with a combination of concentric waves, radial streaks, and negative bowls. For the area that would theoretically benefit the most, around Sgr B2 Z10.24, we see less than 10 percent decrease in noise with source flux remaining the same. 
}

\refreport{
None of the methods described above would change the results of this work or have a meaningful impact on reported fluxes. Other sources of uncertainty, such as self-calibration process or the choice of source extraction method play a more significant role. Thus, in this work we reduce each pointings separately, splitting the overlapping area between the two pointings based on the distance from the center of the pointing and the source clustering.
}

\subsection{Evaluation using unsharp masking}
We attempted to use unsharp masking to evaluate at what value lowering cleaning threshold no longer improves the image on small scales. To create a sharpened image, the original image is first smoothed using a Gaussian kernel. Smoothing removes ``blurs" small scale structures effectively removing them while keeping the large scale structures mostly intact. Different size of the smoothing kernel can be chosen to effect different size-scales. Typically, the sharpened image is:
\begin{equation}
    \mathrm{Sharpened} = \mathrm{Original} \times A - \mathrm{Smoothed} \times (A - 1),
\label{eqn:unsharp}
\end{equation}
where $A$ determines how strong the effect is.

Generally, cleaning to a lower threshold results in ``better" images. However, cleaning too deep results in the \texttt{tclean} model including artifacts. In highly complex, high-dynamic-range images some of the artifacts will be present in the model unless an extremely shallow clean is performed. We strove to achieve a balance between cleaning depth and the amount of artifacts in the image model. We found that decreasing the cleaning threshold always decreases the measured RMS, which consequently could not be used to determine the ``best" cleaning threshold.

The majority of the artifacts in our images are small-scale: concentric rings and radial streaks with scale-length smaller than the beam size. We used a simplified version of unsharp masking, $\mathrm{Sharpened} = \mathrm{Original} - \mathrm{Smoothed}$, to create large-scale-structure-free versions of our images at different cleaning thresholds. If there exists a cleaning threshold at which small-scale artifacts are no longer removed by cleaning, we should be able to detect it by measuring the RMS of these sharpened images. However, we found no such correlation and the measured noise in sharpened images steadily decreased with the cleaning depth.

\begin{figure}[ht]
    \centering
    \fig{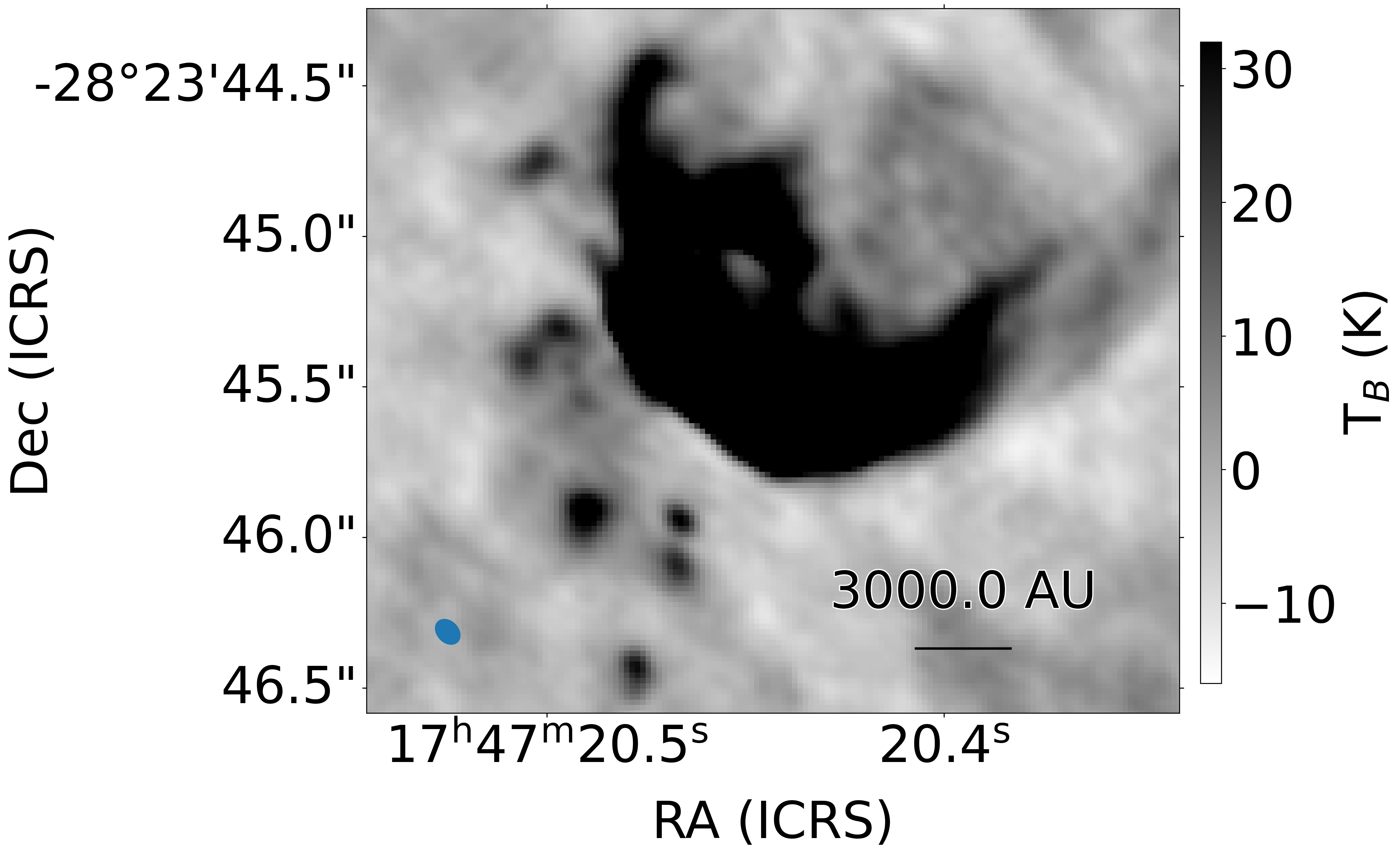}{0.48\textwidth}{(a) Final calibrated image of Sgr B2 S \hii region and surrounding cores.}
    \fig{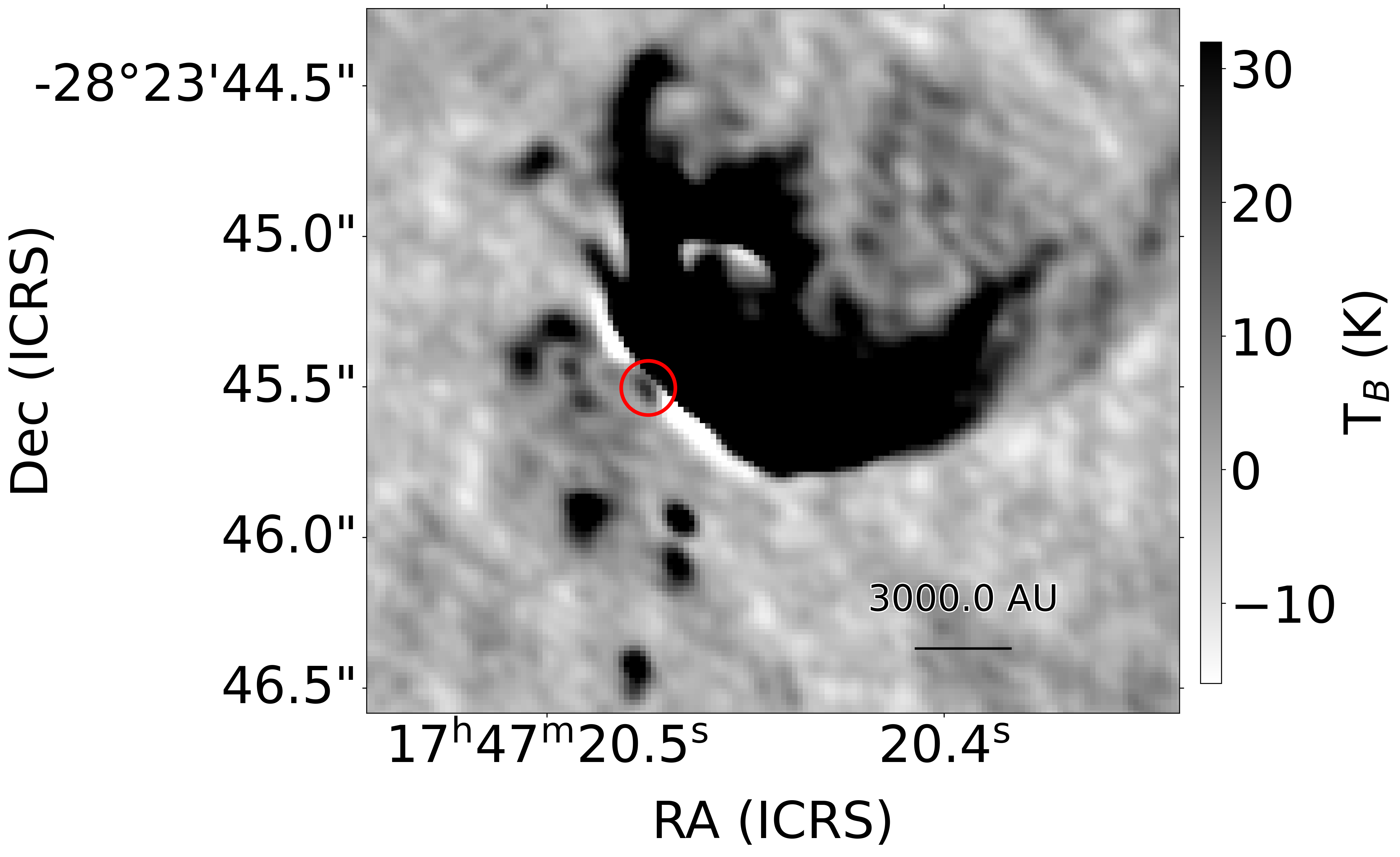}{0.48\textwidth}{(b) The same region with unsharp mask applied. A new source is now visible very close to the \hii region, circled in red. 
    This source blends with the \hii region in the original region and cannot be picked up by the source extraction algorithms. Thus, this source is not included in our catalog.}

\caption{Unsharp masking revealing a source very close to a bright \hii region.}
\label{fig:unsharp_masking}
\end{figure}

During the unsharp masking investigation we made sharpened versions of the final images. Upon inspection, we identified a core-like source close to Sgr B2 S \hii region as shown in Figure \ref{fig:unsharp_masking}. Because of the core's proximity to the bright \hii region, the chosen source extraction parameters are not able to identify it as a separate source. Thus, this source is not included in our final catalog. While unsharp masking is a viable method to identify such otherwise ``hidden" sources, the sharpened images cannot be used for data analysis as the flux is not conserved during the transformation.

\section{Source catalog}
We present an excerpt from Sgr B2 N core catalog at 3 mm in Table \ref{tab:catalog}. The full catalogs are available in machine-readable format and on GitHub: \url{https://github.com/nbudaiev/SgrB2_ALMA_continuum}.

\begin{deluxetable}{ccccccccccccccc}
\rotate
\tabcolsep=0.11cm
\tablecaption{First 25 entries of the Sgr B2 N catalog at 1 mm. The full catalogs for each pointing are available in machine-readable format. \label{tab:catalog}}
\tablehead{
ID & $S_{\nu,flux}$ & $S_{\nu,peak}$ & $S_{\nu, err}$ & FWHM$_{maj}$ & FWHM$_{min}$ & RA (ICRS) & Dec (ICRS) & Score & Source & ID$_{\mathrm{Band 3}}$ & $T_B$ & $\alpha$ & $\alpha_{err}$ & Mass \\
 & $\mathrm{mJy}$ & $\mathrm{mJy\,beam^{-1}}$ & $\mathrm{mJy\,beam^{-1}}$ & $\mathrm{{}^{\prime\prime}}$ & $\mathrm{{}^{\prime\prime}}$ &  &  &  &  &  & $\mathrm{K}$ &  &  & $\mathrm{M_{\odot}}$}
 
\startdata
1 & 13.4 & 7.2 & 0.353 & 0.11 & 0.043 & 17:47:19.9296 & -28:22:19.272 & 2 & core & 6 & 53 & 2.3 & 0.1 & 7.5 \\
4 & 17.9 & 5.2 & 0.353 & 0.307 & 0.048 & 17:47:19.9608 & -28:22:18.624 & 1 & core & 11 & 39 & 1.5 & 0.2 & 10.4 \\
7 & 5.6 & 11.9 & 0.353 & 0.046 & 0.018 & 17:47:19.8744 & -28:22:18.588 & 1 & HII\_cand & 14 & 89 & -1.01 & 0.05 &  \\
8 & 76.7 & 16.5 & 0.353 & 0.134 & 0.087 & 17:47:19.884 & -28:22:18.3 & 1 & HII\_cand & 17 & 123 & 1.08 & 0.03 &  \\
11 & 11.7 & 6.1 & 0.353 & 0.065 & 0.064 & 17:47:19.9464 & -28:22:19.956 & 2 & core & 2 & 45 & 2.18 & 0.08 & 6.7 \\
15 & 11.3 & 6.3 & 0.353 & 0.078 & 0.041 & 17:47:19.9368 & -28:22:19.812 & 2 & core & 3 & 47 & 2.5 & 0.08 & 6.4 \\
23 & 7.5 & 6.0 & 0.353 & 0.064 & 0.036 & 17:47:19.9152 & -28:22:19.632 & 1 & core &  & 45 & 2.82 & 0.09 & 4.3 \\
25 & 6.7 & 5.6 & 0.353 & 0.056 & 0.039 & 17:47:19.9056 & -28:22:19.596 & 1 & core &  & 42 & 2.66 & 0.09 & 3.9 \\
26 & 17.3 & 6.8 & 0.353 & 0.103 & 0.049 & 17:47:19.8936 & -28:22:19.524 & 1 & core &  & 51 & 1.8 & 0.07 & 9.7 \\
27 & 7.4 & 5.1 & 0.353 & 0.066 & 0.04 & 17:47:19.8744 & -28:22:19.524 & 1 & core &  & 38 & 1.41 & 0.09 & 4.3 \\
28 & 5.5 & 8.0 & 0.353 & 0.044 & 0.027 & 17:47:19.92 & -28:22:19.56 & 1 & core &  & 59 & 2.68 & 0.06 & 3.1 \\
31 & 20.9 & 12.9 & 0.353 & 0.091 & 0.033 & 17:47:19.9104 & -28:22:19.452 & 1 & core & 4 & 96 & 2.0 & 0.04 & 11.3 \\
55 & 10.7 & 6.8 & 0.353 & 0.12 & 0.031 & 17:47:19.8336 & -28:22:18.3 & 2 & core & 18 & 50 & 1.9 & 0.09 & 6.0 \\
59 & 51.6 & 8.4 & 0.353 & 0.2 & 0.077 & 17:47:19.836 & -28:22:18.012 & 1 & core & 22 & 62 & 2.53 & 0.06 & 28.6 \\
64 & 8.2 & 8.9 & 0.353 & 0.05 & 0.031 & 17:47:19.8888 & -28:22:17.94 & 1 & core &  & 66 & 1.57 & 0.05 & 4.5 \\
68 & 4.2 & 5.0 & 0.353 & 0.054 & 0.027 & 17:47:19.8456 & -28:22:17.904 & 1 & core & 26 & 37 & 2.1 & 0.1 & 2.5 \\
71 & 10.1 & 9.4 & 0.353 & 0.052 & 0.035 & 17:47:19.8864 & -28:22:17.832 & 1 & core &  & 70 & 1.27 & 0.04 & 5.5 \\
74 & 5.2 & 4.2 & 0.353 & 0.055 & 0.046 & 17:47:19.9056 & -28:22:17.76 & 1 & core & 29 & 31 & 2.5 & 0.1 & 3.2 \\
75 & 35.1 & 8.9 & 0.353 & 0.115 & 0.067 & 17:47:19.872 & -28:22:17.688 & 1 & HII\_new & 30 & 66 & 1.48 & 0.05 &  \\
77 & 4.6 & 5.8 & 0.353 & 0.048 & 0.03 & 17:47:19.9416 & -28:22:17.724 & 1 & core & 27 & 43 & 2.16 & 0.08 & 2.6 \\
81 & 12.2 & 9.8 & 0.353 & 0.09 & 0.027 & 17:47:19.98 & -28:22:17.58 & 1 & core & 31 & 73 & 2.06 & 0.05 & 6.7 \\
85 & 20.2 & 7.9 & 0.353 & 0.088 & 0.058 & 17:47:19.872 & -28:22:17.472 & 2 & core & 33 & 59 & 2.59 & 0.06 & 11.2 \\
86 & 18.7 & 5.9 & 0.353 & 0.122 & 0.062 & 17:47:19.8936 & -28:22:17.472 & 1 & core &  & 44 & 2.52 & 0.09 & 10.7 \\
87 & 40.5 & 13.7 & 0.353 & 0.106 & 0.056 & 17:47:19.9872 & -28:22:17.436 & 1 & core & 32 & 102 & 1.89 & 0.03 & 21.8 \\
88 & 35.5 & 6.3 & 0.353 & 0.148 & 0.1 & 17:47:19.9392 & -28:22:17.328 & 1 & core & 35 & 47 & 1.91 & 0.08 & 20.2 \\
\enddata
\end{deluxetable}

\section{Radial core distribution}
\begin{figure}[ht]
    \centering
    \includegraphics[width=0.5\columnwidth]{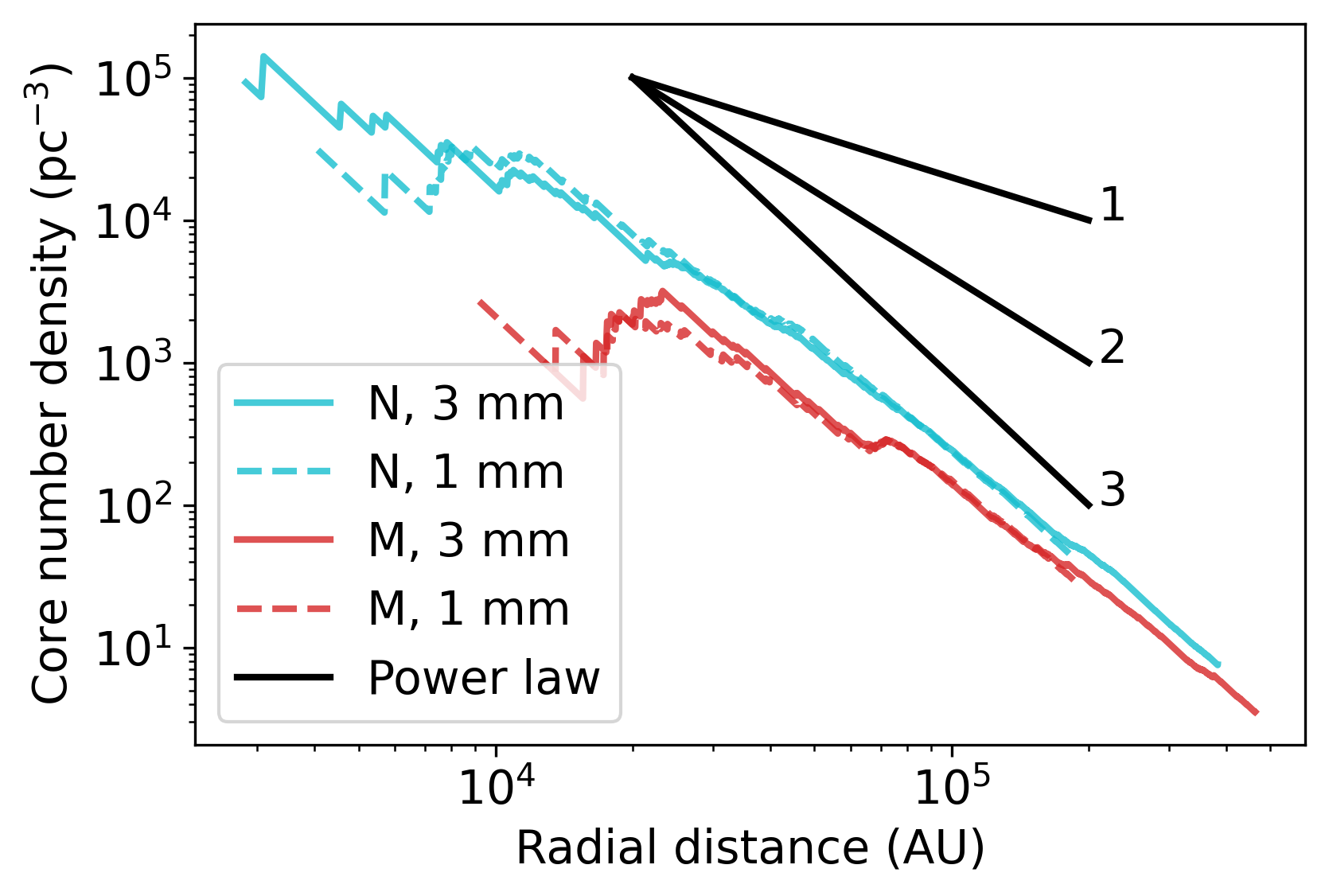}
    \caption{Radial core distribution as a function of number density. Several power law slopes are overplotted. The slope of the core distribution varies from $\sim$2 near the center of the clusters to $\sim$3 at the edges. The small bump at $7\times10^4$ AU for Sgr B2 M is due to a large number of cores around Sgr B2 South.}
    \label{radial_distribution}
\end{figure}

We show the core volume number density distribution as a function of distance from the center of each of the two main clusters in Figure \ref{radial_distribution}. The distributions follow a power law with a slope of 2 near the center and increases to 3 at the very edges. This distribution calculation does not take into account the radial decrease in sensitivity of our data, thus the core number density at large distances is underestimated. A sudden increase in the core number density for Sgr B2 M at $7\times10^4$ AU is caused by a large number of sources in the vicinity of Sgr B2 South.

\end{document}